# Molehills into mountains: Transitional pressures from household PV-battery adoption under flat retail and feed-in tariffs

Kelvin Say[a,b*], Michele John[b]

[a] *Energy Transition Hub, University of Melbourne, Grattan Street, Parkville 3010, Australia*
[b] *Sustainable Engineering Group, Curtin University, Kent Street, Bentley 6102, Australia*
[*] *Corresponding author: kelvin.say@unimelb.edu.au (K. Say)*

## HIGHLIGHTS

- Flat retail tariffs and FiTs affect PV-battery investments and reshape grid-utilisation
- Annual grid-imports significantly reduced, grid-exports continue to increase
- Residual household demand becomes increasingly winter dominant
- Residual household demand drives the emergence of early-morning diurnal peak demand
- Role of the retailer shifts as PV-battery households begin to export more than consumed

## ABSTRACT

With Australia's significant capacity of household PV, decreasing battery costs may lead to widespread use of household PV-battery systems. As the adoption of these systems are heavily influenced by retail tariffs, this paper analyses the effects of flat retail tariffs with households free to invest in PV battery systems. Using Perth, Australia for context, an open-source model is used to simulate household PV battery investments over a 20-year period. We find that flat usage and feed-in tariffs lead to distinct residual demand patterns as households' transition from PV-only to PV-battery systems. Analysing these patterns qualitatively from the bottom-up, we identify tipping point transitions that may challenge future electricity system management, market participation and energy policy. The continued use of flat tariffs incentivises PV-battery households to maximise self-consumption, which reduces annual grid-imports, increases annual grid-exports, and shifts residual demand towards winter. Diurnal and seasonal demand patterns continue to change as PV-battery households eventually become net-generators. Unmanaged, these bottom-up changes may complicate energy decarbonisation efforts within centralised electricity markets and suggests that policymakers should prepare for PV-battery households to play a more active role in the energy system.

# 1. Introduction

As set out by the Paris agreement, the decarbonisation of the power sector is necessary to mitigate the effects of global warming. The growth and integration of utility-scale renewable energy technologies has changed the economic and operational dynamics that have traditionally underpinned power sector management and planning. Reductions in the cost of PV technology have also benefited households by reducing cost barriers for customers to self-generate and reduce future electricity bills. Over the past decade, the rapid and widespread rise of household PV in Australia (Australian Photovoltaic Institute, APVI, 2019a, 2019b) has noticeably impacted whole of system operation and wholesale electricity market dynamics (Australian Energy Market Operator, AEMO, 2019b). With lithium-ion battery energy storage costs decreasing (Schmidt et al., 2017) there has been increased use within the utility-scale power sector (International Renewable Energy Agency, IRENA, 2019) and early-adoption at the household-scale (Graham et al., 2019; Porteous et al. 2018; SunWiz, 2020). As widespread household PV-battery adoption has the potential to further erode centralised electricity supply markets (Agnew and Dargusch, 2015), it is necessary to better understand the extent of these changes to the power sector.

The aim of this paper is to assess how maintaining flat retail usage charges and Feed-in Tariffs (FiTs), that influence household investments in PV and battery systems, affect residual demand and influence the interconnected layers of traditional liberalised electricity markets, including its distribution network, market dispatch, and electricity retailing. By identifying transition patterns from household PV battery adoption and their qualitative effects on the power sector, system managers and policymakers may better understand (and prepare for) their wider adoption.

In this paper we analysed changes in aggregate household grid-utilisation within the context of an islanded electricity system and liberalised energy market in Perth, Western Australia, and assessed the power sector challenges and opportunities afforded from the growth of household PV battery adoption. Cumulative household PV capacity (1 $GW_P$ at the end of 2019) significantly exceeds the capacity of utility PV and wind (10 $MW_P$ and 478 MW respectively) (AEMO, 2019d). In a system with 4 GW of peak demand, households have an outsized ability to influence the operational and market layers within this system. We utilised an open-source PV battery investment model (Say et al., 2019) applied to 261 real household underlying demand and insolation profiles. The aggregate grid-utilisation changes (at a half-hour resolution) were used to establish a pattern of diurnal and seasonal transitions across a range of proportional FiTs (valued at a fixed percentage of the retail usage charge). The research focused on the household sector with business-as-usual retail conditions, that maintained flat and increasing retail tariffs, decreasing PV and battery system costs, and a 5 $kW_P$ FiT eligibility limit. As flat tariffs do not value the timing of consumption and exports, grid-charging and grid-discharging from the battery was not evaluated. Reflecting business-as-usual, retail market conditions for the household sector were simulated over a 20-year period to analyse emerging power sector tensions from the continued use of flat retail tariffs. This analysis may also apply more broadly in other regions that use flat retail tariffs and are experiencing growing household PV battery deployment. The provision of the open-source model also contributes to the literature by providing transparency and reproducibility for subsequent research.

In summary this paper reviews the critical impacts that household PV-battery adoption will potentially have on the Western Australian electricity market. The impact of flat retail tariffs on household PV battery investment, and the transition that is forced onto electricity generators in terms of the movement towards early morning diurnal peak demand and a shift towards winter dominant residual demand. This transition results in a number of significant changes in the Western Australian energy market. Firstly, PV-battery households becoming net-generators. Secondly, changes to the diurnal timing and ramping of utility-scale generation. Thirdly, the potential for utility providers to capitalise on under-utilised flexibility from PV-battery households. However, this transition also results in certain tipping points being reached for current grid operators, resulting in falling retailer revenues that increases the opportunity for household DER to provide flexible generation and load through greater energy market participation.

This paper is structured as follows: Section 2 presents the background and literature review. Section 3 describes the modelling and the analytical approach. Section 4 characterises the modelling results into various grid-operation stages and determines the overall transition behaviour. Section 5 discusses the transition patterns that emerge starting from the aggregate household level through to their wider power sector effects, along with the limitations and outlook. Section 6 concludes with key findings and policy implications.

# 2. Background and literature review

## 2.1. The growth of household PV in Australia

With over 1 $GW_P$ of cumulative PV capacity installed behind-the-meter across 27% of free-standing households (Australian Photovoltaic Institute, APVI, 2019a), the South West Interconnected System (SWIS) in Western Australia has experienced a fundamental shift in how electricity is used over the last decade. Cumulatively, behind-the-meter household PV (or rooftop

PV) capacity has become the largest generator on the network[1], outstripping commercial and industrial customer-sited PV capacity (APVI, 2019b) and has been recorded supplying over 45% of instantaneous network demand (AEMO, 2019b) on the SWIS.[2] Over the next decade (2019 to 2028) the amount of energy generated from rooftop PV is expected to more than double from 1657 GWh to 3432 GWh respectively (AEMO, 2019d). Currently rooftop PV only contributes to less than 10% of the overall energy demand on the SWIS, but new minimum network demand records are being exceeded such that system stability has become an increasing concern (AEMO, 2019a) and mechanisms are being considered that can remotely curtail household generation (AEMO, 2020b) or encourage the use of front-of-the-meter energy storage systems (Mercer, 2019a). As a small- to medium-sized islanded electricity system, the SWIS lacks the ability to export excess generation, which makes it more sensitive to changes in grid-utilisation when compared to larger and interconnected electricity systems. This heightens the need for a better understanding of how future customers with PV battery systems may interact with the grid, while providing context for what larger systems may experience in the future.

The high penetration of rooftop PV is not unique to Western Australia and applies across most Australian states and territories.[3] This has been driven by continued reductions in installed system costs (Solar Choice, 2019a), abundant solar resources (World Bank Group, 2019), relatively high retail electricity tariffs (Australian Energy Market Commission, AEMC, 2018a) with low consumer trust (AEMC, 2018b) and increasing community concern for greenhouse gas mitigation. With behind-the-meter Distributed Energy Resources (DER) being unmonitored and uncontrolled (Australian Energy Market Operator, AEMO, 2019a, 2018), system operation remains highly reactive and sensitive to aggregate changes in household grid-utilisation.

[1] The SWIS has a peak network demand of 4.4 GW. The largest utility-scale generator on the network is the Muja coal power station with a nameplate capacity of 854 MW (AEMO, 2019d).
[2] This was over a 30-min trading interval and occurred on 29 September 2019.
[3] Over 2 million households (or 20% of all free-standing households) having installed rooftop PV systems (APVI, 2019a, 2019b). As of the end of 2018, rooftop PV systems under 10 $kW_P$ accounted for more than half of the nation's cumulative installed PV capacity (APVI, 2019b). Average system capacity continues to rise and exceeded 7 $kW_P$ in 2018 (AEC, 2019).

The declining costs of batteries (Schmidt et al., 2017) may lead to a widespread adoption of household PV-battery systems (Parkinson, 2018) that further change how households utilise and interact with the grid. With household PV-only systems, all behind-the-meter generation across the network use the same solar resource simultaneously, which aligns their temporal effects and leads to observable system level patterns like the *duck curve* (Denholm et al., 2015; Maticka, 2019). With PV-battery systems, an additional degree of freedom is provided by the choice of battery capacity. This choice further depends on financial factors (such as retail tariff structures and feed-in tariffs) and technical factors (such as underlying household demand) that influence the installed level of self-generation and storage capacity, along with how it is dispatched. These dependencies make anticipating changes to grid-utilisation with PV-battery adoption less certain (e.g., retail tariff incentives may influence an existing PV-only household to install a large capacity battery with additional PV capacity or install a smaller capacity battery and maintain the existing PV capacity). As each choice leads to different household grid-utilisation profiles, understanding how retail tariffs influence this process (and consequently the underlying layers of the power sector) becomes critically important for the design of retail energy policies and suitable market design.

Though electricity systems remain region specific, similar processes underpin liberalised electricity markets, namely wholesale energy markets that competitively dispatch from the lowest marginal cost generators, regulated transmission and distribution network monopolies and retailers that hedge wholesale prices to provide simpler tariffs to customers. These similarities allow qualitative analyses from the SWIS to apply more broadly to other markets with similar structural designs.

## 2.2. Literature review

As greater amounts of renewable energy generation are incorporated into liberalised electricity markets, operation patterns begin to emerge from interactions between the system and market layers, such as the *merit-order effect*[4] from zero-marginal cost generators (Sensfuß et al., 2008), and the *duck curve*[5] from growing PV generation (Denholm et al., 2015). While each electricity system is unique, similar technical and economic foundations have meant that the *merit-order effect* and *duck curve* have been widely generalisable. By considering households as rational investors that affect the various layers of the electricity system, we are able to build upon the extensive literature on financial investment modelling with renewable energy technologies, and analytical frameworks used to analyse energy transitions.

### *2.2.1. Household PV battery investment modelling*

With Australia's current leadership in behind-the-meter PV adoption (Australian Energy Council, AEC, 2016; APVI, 2019a) and early-stages of PV-battery adoption, there remains insufficient information for an *ex-post* analysis. However, an *ex-ante* financial investment perspective provides a techno-economic foundation to frame future investment decisions that can be useful to evaluate potential futures (Wüstenhagen and Menichetti, 2012). At a household-scale the financial investment perspective allows a range of possible PV battery system capacities to be framed as a set of competing investment opportunities based upon their expected electricity bill savings and upfront costs. Financial metrics using discounted cash flows, such as Net Present Value (NPV), Internal Rate of Return (IRR), and Discounted Payback Period (DPP) are commonly used to assess the value of each investment opportunity. The PV battery configuration with the highest financial value provides an indication of the systems that households may choose to install in the future and the costs necessary to achieve this.

[4] The reduction in wholesale electricity prices that occurs as the capacity of near-zero short run margin cost generation (namely wind and solar PV) increases within an electricity market, due to the merit order and dispatch price determination.

[5] The *duck curve* describes how network operations could be impacted from increasing solar PV capacity. Operationally, minimum demand moves into the middle of the day and gradually declines, resulting in a risk of overgeneration. Furthermore, as the late afternoon peak persists, the ramping required to meet the peak increases for all remaining generators.

Schram et al. (2018) used real utility net-meter data from 79 PV-only households in Amersfoort, Netherlands and determined potential electricity bill savings across a range of battery capacities. With a battery simulation model (that maximised PV self-consumption) under flat retail and feed-in tariffs, the cost-optimal battery capacity for each household was calculated using NPV analysis. Using these cost-optimal battery capacities, they found alternative battery operating modes could significantly reduce winter peak demand and that increasing battery capacities beyond the cost-optimal configuration only slightly reduced overall profitability. This suggests there is an opportunity for joint investment between households and utilities to further improve peak demand reduction. Schopfer et al. (2018) calculated NPV across a range of predetermined PV battery combinations and system costs using real energy consumption data from 4190 households in Zurich, Switzerland with a PV battery simulation model (also maximising self-consumption) and time-of-use retail tariffs and flat FiTs. With 2018 PV and battery system costs of 2000 €/kWP and 1000 €/kWh respectively, PV-only systems were profitable for less than half of the households and PV-battery systems remained unprofitable; however as battery costs decreased to 250 to 500 €/kWh, a tipping point emerged with the majority of households having profitable PV-battery systems. The use of real energy consumption data was the focus of Linssen and Stenzel (2017) that showed aggregate or synthetic data could lead to an overestimation of economic feasibility. Other examples utilising a cost-optimal household PV battery perspective, include Dietrich and Weber (2018), Hoppmann et al. (2014), Talent and Du (2018), von Appen and Braun (2018), and Weniger et al. (2014). Each of these bottom-up PV battery studies however, have used a 'greenfield' (or one-shot investment perspective), while 'brownfield' perspectives that focus on incremental investments to assess the path of cost-optimality are rarely evaluated in household PV battery literature. Real options models have been used previously to assess the effect of different policy conditions on the timing and scale of renewable energy investment decisions at the utility-scale (Reuter et al., 2012), but only recently used to evaluate PV battery investments (Ma et al., 2020).

### *2.2.2. Electricity system transitions*

By being able to consume, self-generate and store energy, customers with PV and battery systems are not passive actors in the electricity system, but rather active participants that react to prices and expectations (Klein and Deissenroth, 2017) with the ability to influence broader energy system transitions. Energy transitions are described as co-evolving relationships between techno-economic, socio-technical, and political perspectives with transition pathways being a series of reconfiguring systems driven by a multitude of competing actors (Bolwig et al., 2019; Cherp et al., 2018; Geels et al., 2017; Pfenninger et al., 2014). Due to the complex interactions between different layers of the energy system, that extend beyond purely numerical assumptions, these studies highlight the importance of using combined qualitative and quantitative analysis to evaluate the energy system transitions. For example, Schill et al. (2017) modelled the implications of direct and indirect support mechanisms with various PV-battery operating strategies. Numerical data was used as a basis for a qualitative evaluation on the potential role of household PV-battery adoption in the German electricity sector. A range of qualitative system-level arguments for and against household PV-battery adoption were established, such as private rather than public capital can be used to increase renewable energy penetration, through to increased data protection amid rising security concerns. The authors also highlighted the potential of reducing system operation costs by encouraging system-friendly household battery operation. Neetzow et al. (2019) analysed various policy mechanisms that incentivise market friendly household PV-battery operation while reducing the need for network capacity expansion. They find that grid feed-in policies should be complemented by load policies to incentivise households to operate PV-battery systems in a system-friendly manner (i.e., utilising the spare capacity of the electricity system, rather than exacerbating its constraints). They caution policymakers that careful policy design is necessary as battery systems can (if unchecked) exacerbate both load and supply issues across the distribution network. Eid et al. (2016b) constructed a framework of various local energy management market designs from European case studies. By evaluating the socio-economic constructs and regulatory environments, they qualitatively discussed the range of changes necessary to integrate DER systems in a system-friendly manner, along with the challenges and opportunities with this transition. By using a combination of quantitative and qualitative approaches, these studies analyse a broader range of energy system integration outcomes and offers further context to be applied in other regions.

### *2.2.3. Modelling approach*

Analysing the interactions between retail tariffs, household installations of PV and battery capacity, aggregate residual demand, the distribution network, wholesale market dynamics, and existing utility-scale generators requires an ever-increasing number of parameters and assumptions (Bale et al, 2015). Many of these elements depend on socio-political factors that cannot be entirely represented numerically (e.g. householders' personal decisions to install PV battery systems, to political pressure to maintain favourable policy mechanisms such as flat feed-in tariffs). This paper addresses this modelling gap by using a range of scenarios with a bottom-up and combined quantitative and qualitative approach. This approach analyses how PV battery investing households (under flat retail and feed-in tariffs) may qualitatively influence various power sector layers. Energy transition pathways for individual households are generated by considering PV battery adoption as a series of discrete and incremental investment opportunities. This brownfield approach focuses on how household PV battery investment pathways change over time, and how they can lead to aggregate changes in grid-utilisation. With different grid-operation stages emerging, their transitions form a basis to qualitatively assess (i.e. describe the important inter-relationships and dependencies) the power sector impacts. To the best of the authors' knowledge, no studies to date have modelled and characterised how decreasing PV and battery costs combined with the continued use of flat retail and feed-in tariffs leads to transitional pressures from household residual demand on the power sector's system operation, market and retail energy policies.

# 3. Methodology and case study

## 3.1. Analytical framework

This study uses a techno-economic investment simulation model used in previous studies (Say et al., 2019, 2020) called Electroscape. The model considers the range of PV and battery configurations available to a household as a set of competing investment opportunities (based on electricity bill savings). By utilising an investment decision tree (based on real options evaluation), projections of annual PV battery installed capacities are simulated across a range of households using their own underlying demand and insolation profiles. This numerical model establishes how household grid-utilisation may change under (exogeneous) retail market conditions. By categorising and framing these changes as a series of grid-operation transition stages, the qualitative effects on the power sector is evaluated.

With a focus on flat retail tariffs, this paper evaluates five scenarios that vary the relative value of the FiT with respect to the retail usage tariff (volumetric), along with high and low growth scenarios in the sensitivity analysis (Appendix B). FiT payments are only applied to the amount of energy exported after first being consumed by underlying demand. Using FiT conditions that are representative of flat FiTs in Australia (AEC, 2018; AEMC, 2018a) and abroad,[6] five FiT scenarios are modelled that correspond to setting the FiT between 0% and 100% of the retail usage charge (using steps of 25%) and are only eligible for households with PV systems 5 $kW_P$ and under. These five FiT scenarios are named $FiT_0$, $FiT_{25}$, $FiT_{50}$, $FiT_{75}$ and $FiT_{100}$, and value the FiT respectively at 0%, 25%, 50%, 75% and 100% of the retail usage charge. Therefore, in the $FiT_0$ scenario households are not paid for excess energy exports. In the remaining FiT scenarios, the value of energy exports increases but only applies to households with PV systems 5 $kW_P$ and under (AEMC, 2018a; Solar Choice, 2019c). By independently simulating the five scenarios, we broadly capture situations where FiTs change over time. For example, if the FiT is initially valued at 50% of the retail usage tariff (i.e., \$0.20/kWh for a \$0.40/kWh usage charge) and then gradually reduces to 25% of an increasing retail usage tariff over the next 10 years (i.e., \$0.125/kWh for a \$0.50/kWh usage charge), then the transitional impacts are likely to reside between simulation results from the $FiT_{50}$ and $FiT_{25}$ scenarios.

[6] Germany (Engelken et al., 2018), United Kingdom (Pearce and Slade, 2018), Japan (Kobashi et al., 2020)

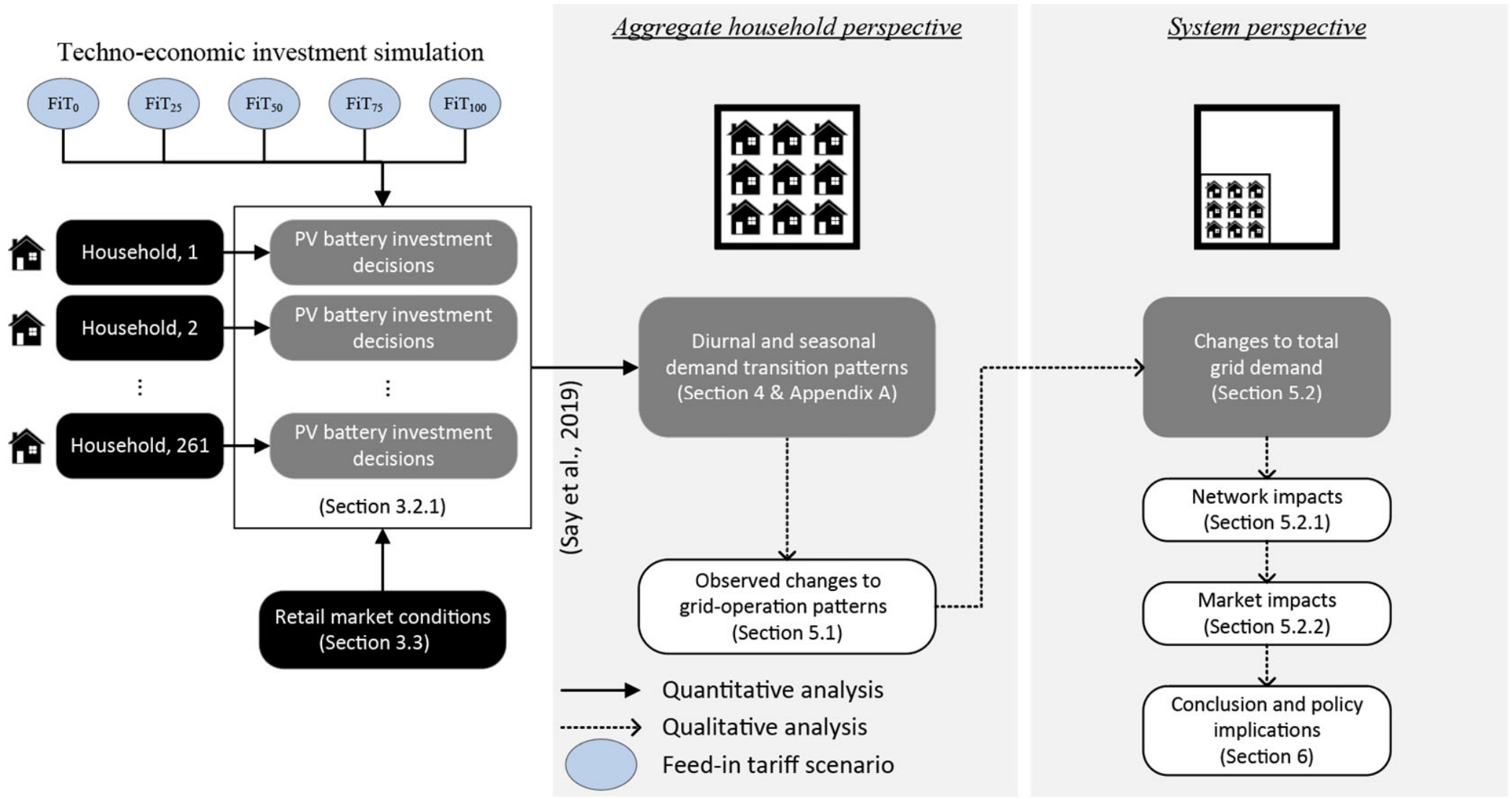

**Fig. 1.** The detailed analytical framework, components, and relationships between the quantitative and qualitative analyses.

*Publishing notes:*
*Displayed over full width*
*Vector graphic provided in "Figure 1.eps"*

Based on business-as-usual conditions, this paper's analysis assumes that PV and battery system prices continue to decrease, and retail usage charges continue to increase. The first part of the analysis (Fig. 1) establishes how different FiT values and flat tariffs, shape and affect household grid-utilisation from household PV battery investments over 20 years. Using a set of representative households, aggregate changes in residual demand affects a range of operational parameters (e.g. the timing and magnitude of annual peak demand) that are then used to quantify how household grid-utilisation changes over time in each FiT scenario. In the second part of the analysis (Fig. 1), common patterns between these grid-utilisation changes are characterised into a set of representative grid-operation transition stages. The trajectory of these transitions provides the foundation for a qualitative assessment on how growing PV battery households may place bottom-up pressure on the power sector's system and market layers. Together the two-part analytical framework (Fig. 2) assesses the policy implications on the power sector from the continued the use of flat retail tariffs. By evaluating market effects from the bottom-up, this paper identifies areas of weaknesses in traditionally centralised liberalised electricity markets and the limitations of using flat retail and feed-in tariffs to manage households' grid-utilisation.

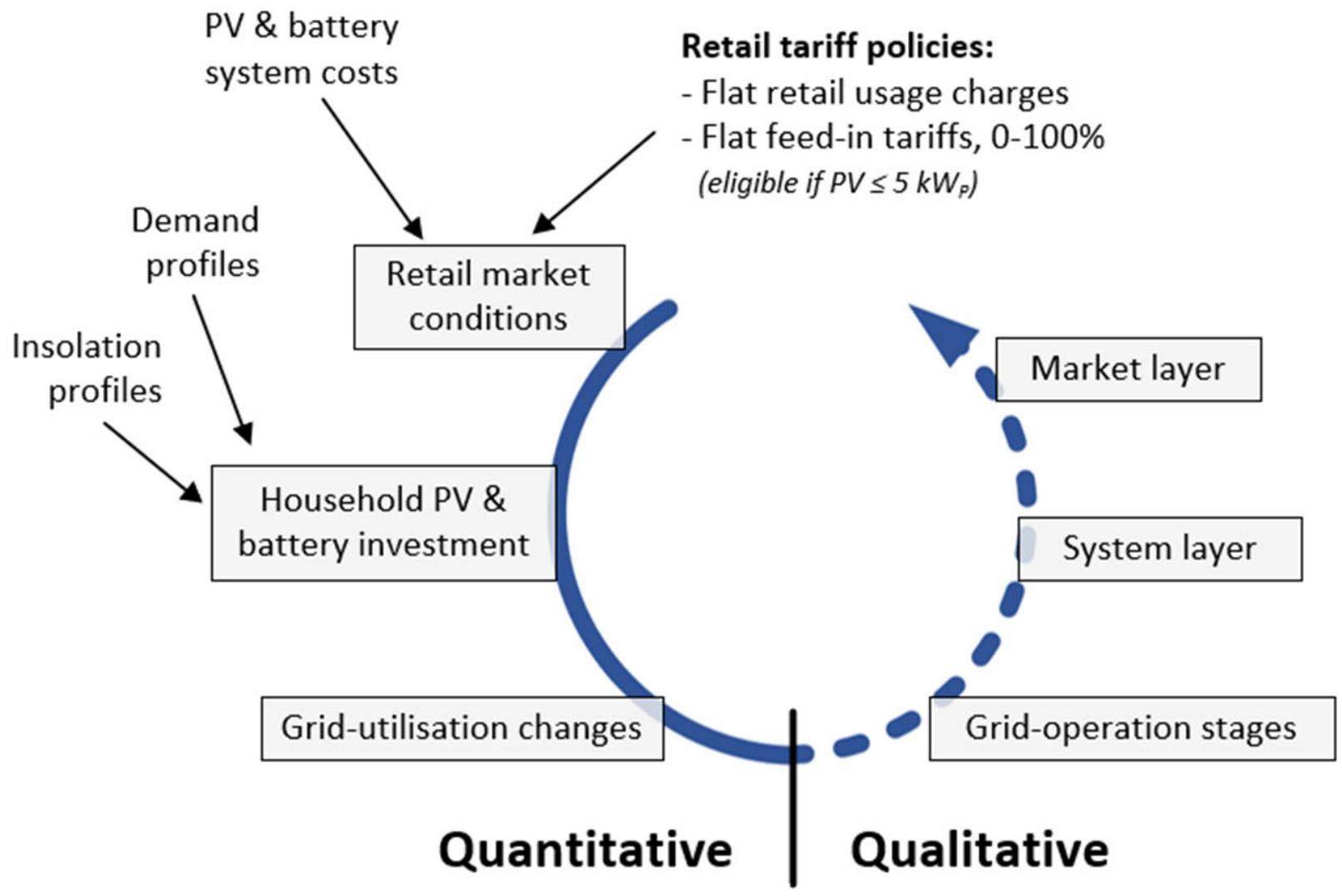

**Fig. 2.** The overall bottom-up analytical framework.

*Publishing notes:*
*Displayed over one column*
*Vector graphic provided in "Figure 2.eps"*

## 3.2. Modelling PV battery adoption

### *3.2.1. Household PV battery investment decision model*

Electroscape is a techno-economic simulation model used to model the timing and capacity of household investments into PV battery systems between 2018 and 2037. Electroscape simulates the investment decisions for each household annually using a 10-year financial horizon. In each year of the simulation, electricity tariffs increase, PV battery system costs decrease, and each household calculates the expected bill savings from installing a range of additional PV and/or battery combinations.[7] As underlying household demand remains constant, these retail market conditions are the sole driver of adoption. NPV of each PV battery combination is calculated by considering expected bill savings over the next 10-years as the *investment cash flow*, system installation costs as *upfront capital costs* and using the average owner-occupied standard variable home loan interest rate as the *discount rate*. This allows each PV battery configuration to be considered as competing investment opportunities. If the financial returns are sufficient to *reduce the uncertainty risk* of making an investment (i.e., requiring a shorter discounted payback period of less than 5 years), the system configuration with the highest NPV is chosen and installed in the given simulation year. This updates the household's future grid-utilisation and subsequent PV battery investments in later years must consider the installed system configuration.[8] Additional detail on the financial and investment modelling assumptions and equations are provided in Appendix C.

This iterative approach allows new PV battery investments to dynamically respond to changing retail conditions while considering previous investments. As the model applies the same investment methodology to each household, variations in installed PV battery system capacities between households are driven by differences in underlying demand and solar insolation profiles. At the end of the simulation, each household produces a half-hourly resolution grid-utilisation profile over 20 years. By aggregating the grid-utilisation across all simulated households, a representation of the grid-utilisation changes at the distribution network level is generated.

---

[7] This is determined by simulating the technical operation from installing additional PV and/or battery system on top of the household's expected grid-utilisation (that considers previously installed PV and/or battery systems). The simulated PV and battery models consider performance degradation, finite operational lifespans, system losses and capacity limits. The resulting differences in annual grid-imports and grid-exports are then valued using the electricity tariffs and FiTs to determine the expected bill savings. More detail is provided in Appendix C.

[8] This sequential investment approach models 'brownfield' investments and allows the economics around the retrofit of existing systems to be modelled explicitly.

### *3.2.2. Technical PV and battery simulation model*

The technical modelling of PV and battery operation follows the simulation framework described by Hoppmann et al. (2014) and uses an AC coupled PV-battery system residing behind-the-meter. The purpose of the technical model is to evaluate, with respect to a household's unique underlying demand and insolation profile, the effect of different PV and battery capacity combinations on a household's grid-utilisation profile over the 10-year financial horizon, which establishes it potential electricity bill savings. The PV generation profile is derived by scaling the household's (kWh/kWP) insolation profile with the PV capacity. By subtracting the PV generation profile from the household demand an intermediate net-demand profile is calculated. The battery simulation model uses this intermediate net-demand profile within its battery capacity constraints to determine the resulting residual demand profile. Grid-charging and grid-discharging is not modelled, as the time-invariant tariffs remove the financial incentive of energy arbitrage. Therefore, the battery dispatch algorithm only aims to maximise PV self-consumption, by charging with excess PV generation until full and discharging to avoid grid-imports until empty (and remaining within the battery inverter limits). Additional detail on the technical modelling assumptions are provided in Appendix C.

## 3.3. Case study parameters and conditions for Perth, Australia

The largest load centre on the SWIS network is the region encompassing the state capital of Perth, Australia. The islanded SWIS network and its liberalised electricity market provides an ideal case study, as it naturally limits the number of external factors. The SWIS has approximately 1.1 million customers with residential, commercial, and industrial sectors consuming 27%, 55% and 18% respectively of its 18 TWh of total annual energy demand (AEMO, 2019d). The vast majority of installed PV capacity resides behind-the-meter on owner-occupied households and the independent market operator continues to expect the household sector to remain as the predominant source of PV capacity growth in the SWIS (Energy Transformation Taskforce, 2020; Graham et al., 2019).

A set of retail market conditions are chosen to reflect business-as-usual conditions in Perth, Australia (Table 1). Household demand and PV generation data was sourced from 261 real households. As disaggregated underlying household demand and insolation profiles are not measured by the utility net-energy meters in Perth, Australia or the wider SWIS network, comparable data[9] from 261 households in Sydney, Australia are used in its place. The dataset (Ausgrid, 2018) was obtained via utility gross-energy meters that separately measured underlying household demand and PV generation. The Sydney dataset is used to represent Perth households as both regions share similar latitudes, climatic conditions, annual energy consumption and solar resources.[10] In addition, household demographics (ABS, 2017a, 2017b) between Perth and Sydney are comparable (average household sizes of 2.6 and 2.8 people respectively) along with median weekly incomes ($1643 and $1750 per household respectively). While Sydney has a lower proportion of owner-occupied dwellings (64%) compared to Perth (73%), the Sydney utility meter data was obtained from owner-occupied and free-standing households (Ausgrid, 2018), reflecting the housing demographic in Perth most likely to invest in PV battery systems (APVI, 2019b). Due to these similarities the Sydney dataset is used to represent the underlying demand and PV generation of Perth households. The aggregate characteristics of the underlying demand data is provided in Appendix A.1.

---

[9] Half-hourly timeseries data was obtained from 300 gross-metered PV households in Sydney, Australia between 1st July 2012 and 31st June 2013. After removing households with missing timeseries data, 261 households remain. The insolation profile (kWh/kW$_P$) for each household was obtained by normalising the solar PV generation profile by their declared PV capacity. Further information on collection of the dataset is documented by Ratnam et al. (2017).

[10] From the Sydney data set, the average annual energy demand per household is 5.62 MWh and average PV capacity factor is 14.8%. This is consistent with Perth that has an average annual energy demand per household of 5.83 MWh (ABS, 2013) and average PV capacity factor of 14.1% (NREL, 2018).

Reflecting SWIS retail conditions, a two-part retail electricity tariff structure is used with an initial usage charge of AU$0.27/kWh (Infinite Energy, 2017) and fixed daily charge of AU$0.95/day. Using historical electricity price increases between 2008 and 2018 (ABS, 2018), retail tariffs are assumed to increase at a fixed rate of 5% per annum.[11] The FiT payments reflect the incumbent retailer conditions (Synergy, 2017) and are only applicable for household PV capacities 5 kWP and under. PV system costs start at AU$1400/kWP[12] (Solar Choice, 2019b) and decrease at -5.9% per annum (Ardani et al., 2018). Linear degradation of PV generation was modelled, with 80% remaining at the end of a 25-year operational lifespan. Battery system costs start at AU$900/kWh[13] (Solar Choice, 2018; Tesla, 2018) and decrease at -8% per annum (BNEF, 2019b). Technical specifications of the battery model are based on currently available residential lithium-ion battery systems, such as the Tesla Powerwall 2[14] and sonnenBatterie.[15] Batteries are simulated with a 100% depth-of-discharge, 5 kW charge and discharge limit, round-trip efficiency of 92% (reflecting warranted performance) and assumes a linear degradation with 70% energy storage capacity remaining at the end of a 10-year operational lifespan. As flat retail tariffs are used, grid-charging and grid-discharging is not simulated. By assuming that households finance the cost of investments using their home loan, a discount rate of 6% is applied, reflecting the 10-year historical average of Australian owner-occupied standard variable home loans (Reserve Bank of Australia, RBA, 2018).

---

[11] This is a simplified assumption that is used to illustrate the effect on household PV battery investments. The low and high sensitivity analyses respectively evaluate inflation rates of 2% and 8% per annum. Significant uncertainty still remains with the trajectory of future electricity prices, with Western Australian wholesale electricity and network costs respectively contributing to approximately 40% and 45% of usage charges (AEMC, 2018a). The attribution of costs between these two components, in a rapidly changing policy and economic environment makes predictions difficult. We therefore utilise the simplified parameter and sensitivity analysis to bound the results within an analysis envelope.

[12] These PV system costs includes the small-scale technology certificate that provides an upfront capital subsidy as part of the federal Renewable Energy Target policy.

[13] As no uniform support mechanisms are currently in place these battery system costs do not include any subsidies.

[14] https://www.tesla.com/en_AU/powerwall

[15] https://sonnen.com.au/sonnenbatterie/

**Table 1**
Input parameters and data used in the study.

| Input Parameter | Unit | Values | Derived from |
|---|---|---|---|
| Scenario forecast period | years | 20 | Model assumption |
| Simulation time step | minutes | 30 | Model assumption |
| Initial flat FiT rebate | AUD / kWh | 0 – 0.27 | Model assumption |
| Initial flat usage charge | AUD / kWh | 0.27 | (Infinite Energy, 2017) |
| Change in tariff charges/rebates | % / a | 5 | (ABS, 2018) |
| FiT rebate installed capacity limit | $kW_P$ | 5 | (Synergy, 2017) |
| Discount rate | % / a | 6 | (RBA, 2018) |
| Initial installed PV system cost | AUD / $kW_P$ | 1400 | (Solar Choice, 2019a) |
| Initial installed battery system cost | AUD / kWh | 900 | (Tesla, 2018) |
| Change in installed PV system costs | % / a | -5.9 | (Ardani et al., 2018) |
| Change in installed battery system costs | % / a | -8 | (BNEF, 2019b) |
| Number of households | household | 261 | (Ausgrid, 2018) |
| Solar PV generation profile (per household) | Wh | Time series | (Ausgrid, 2018) |
| Underlying demand profile (per household) | Wh | Time series | (Ausgrid, 2018) |

<u>*Publishing notes:*</u>
*Preferably display over a single column; if it cannot fit, then over the full width*
*Table provided in “Tables_landscape.docx”*

These case study parameters (Fig. 3) reflect constantly increasing retail tariffs, decreasing PV and battery system costs and an adherence to the existing two-part tariff structure over 20-years. The highly interconnected nature of the electricity market means that growing household PV battery adoption would likely drive further structural and financial changes (such as, new retail tariff structures, reducing wholesale energy costs with greater zero-marginal cost renewable generation, distribution and transmission network upgrades, new decentralised energy markets) that have implications on the future value of DER. Explicitly modelling these future power sector reactions remains outside the scope of analysis. Rather we focus on the current business-as-usual expectations to assess the layers of the power sector susceptible to growing household PV battery adoption, thus highlighting to policy makers the parts of the electricity market that may require further energy policy reform.

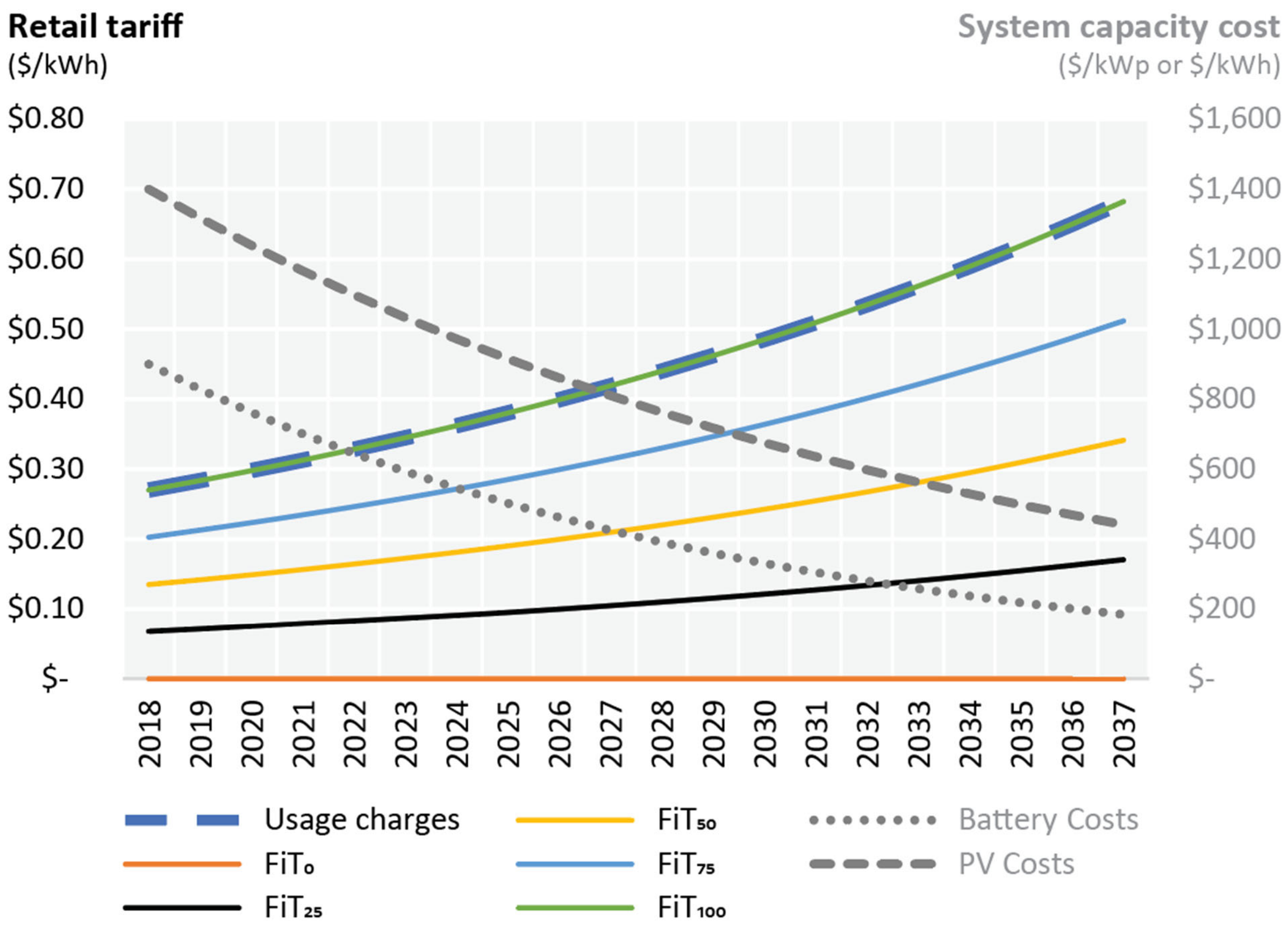

**Fig. 3.** Changes in retail market conditions over the 20-years of the case study.

*Publishing notes:*
*Display over one column*
*Vector graphic provided in "Figure 3.eps"*

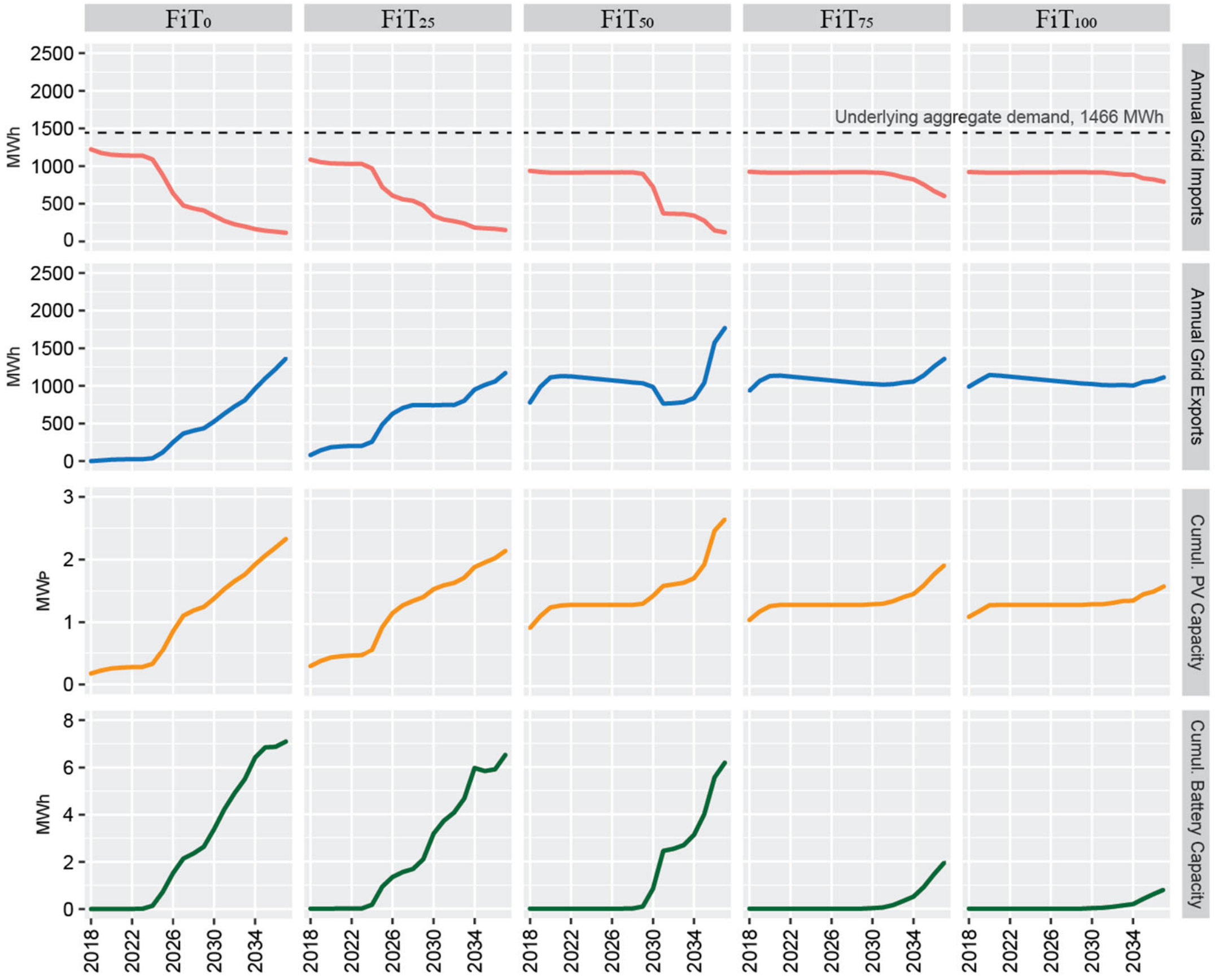

**Fig. 4.** Comparison of operational results (aggregate of 261 households at an annual resolution) projected over 20-years. The feed-in tariff scenarios ($FiT_0$, $FiT_{25}$, $FiT_{50}$, $FiT_{75}$ and $FiT_{100}$) respectively value the FiT at 0%, 25%, 50%, 75% and 100% of the retail usage tariff. FiTs are only eligible for households with PV systems 5 $kW_P$ and under.

<u>*Publishing notes:*</u>
*Display over the full width*
*Vector graphic provided in "Figure 4.eps"*

# 4. Results

Across all five FiT scenarios, with changing retail market conditions (Fig. 3) and the continued use of two-part (and time-invariant) retail tariffs, progressive investments by households eventually lead to PV-battery systems becoming more cost-effective than PV-only systems (Fig. 4). The proportional value of the FiT however influences the timing and magnitude of this transition. With higher FiT scenarios, the installation of PV-only systems begins earlier (and at a higher average capacity) than the lower FiT scenarios, and eventually plateaus at the 5 $kW_P$ per household FiT eligibility limit. Investments in battery capacity occurs, but later than the lower FiT scenarios. In all FiT scenarios, increases in battery storage capacity also coincides with additional PV capacity which indicates that the arrival of cost-effective batteries drives further growth in installed PV capacity. Furthermore, as households increasingly install PV-battery systems, there is an accelerated reduction in annual grid-imports while annual grid-exports continues to increase. This indicates that, under the assumed retail market conditions, households do not find it cost-effective to install battery capacity such that all PV generation is self-consumed. These overall PV battery adoption patterns are a result of differences in investment behaviour, driven by the use of flat usage charges and FiTs.

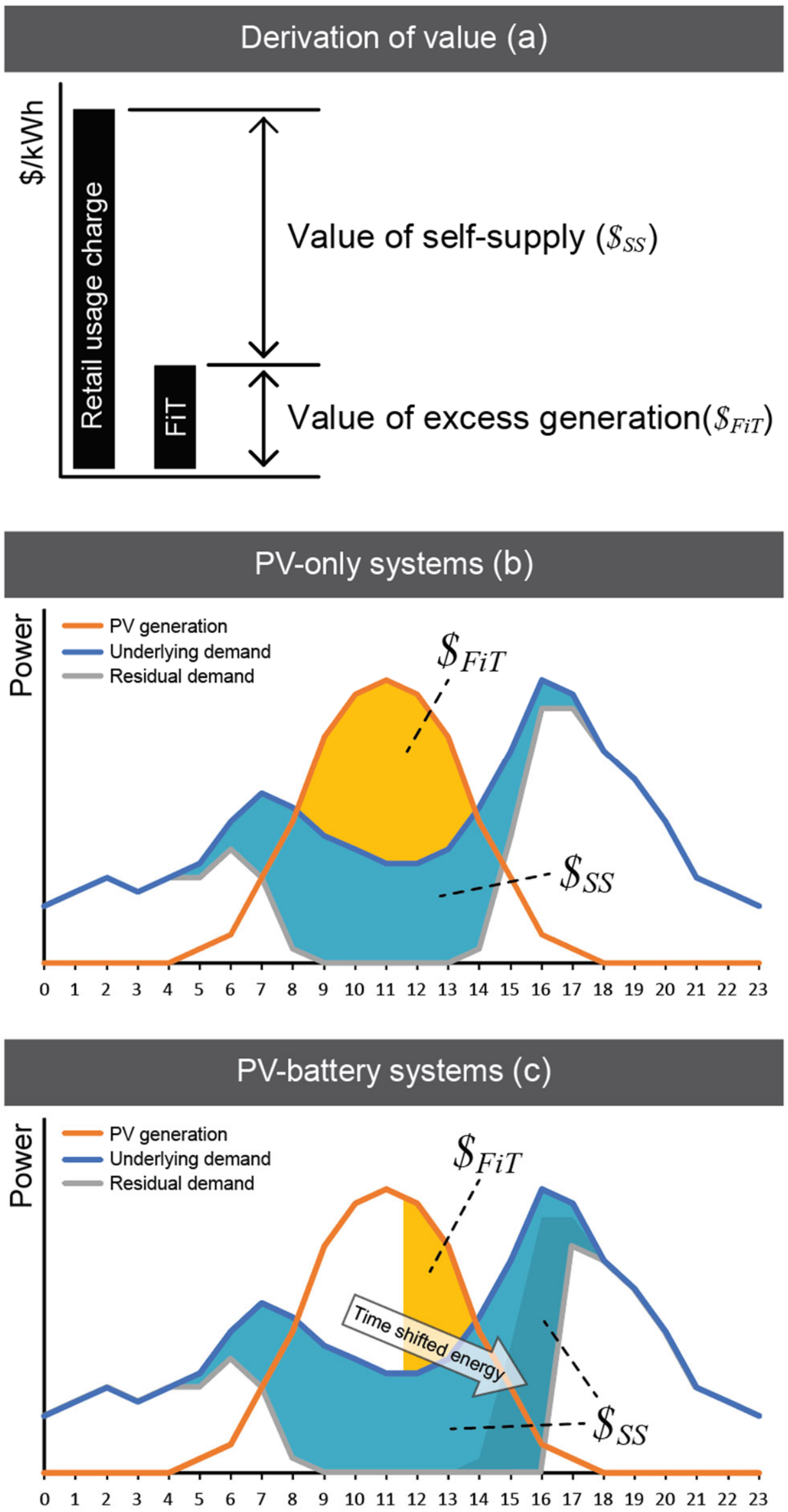

**Fig. 5.** Value stream (stylised) from household self-generation and energy storage using a utility net-energy meter. (a) Derivation of value. (b) Remuneration from PV-only systems. (c) Remuneration from PV-battery systems.

Publishing notes:
Display over a single column.
Vector graphic provided in “ Figure 5.eps”

Under flat tariff structures, PV battery systems offer households two revenue streams (Fig. 5), a value of self-supply ($\$_{SS}$) and a value of excess generation ($\$_{FiT}$). As these are derived from the difference between the retail usage charge and FiT (Fig. 5a): higher FiTs increase the value of excess generation and decrease the value of self-supply (i.e., prioritising grid-exports);

lower FiTs decrease the value of excess generation and increase the value of self-supply (i.e., prioritising self-consumption). With PV-only systems, excess energy is valued at $\$_{FiT}$ while self-consumed generation is valued at $\$_{SS}$ (Fig. 5b). With PV-battery systems however, the amount of energy time-shifted is revalued from the $\$_{FiT}$ to $\$_{SS}$ (Fig. 5c) minus round-trip efficiency losses. Furthermore, the 5 $kW_P$ FiT eligibility limit disincentivises increasing PV generation beyond 5 $kW_P$ and limits the ability of high consumption households to reduce their overall grid-imports. Under each of these FiT scenarios, the various price signals interact and incentivise different investment patterns in the short-term, but collectively begin to follow a similar trajectory over the long-term.

**In low-FiT scenarios** ($FiT_0$ and $FiT_{25}$), the value of self-supply is greater than the FiT (i.e., $\$_{SS} > \$_{FiT}$), hence households are incentivised to dimension their systems to maximise self-consumption while minimising excess PV generation (with respect to overall system costs). During the time that battery systems remain cost-prohibitive a lower average PV capacity is installed, as low FiTs disincentivise the installation of excessively large PV systems due to decreasing marginal benefits with increasing PV capacity. However, once battery systems become cost-effective, households have an option to either, (i) size the battery capacity to utilise existing PV generation, or (ii) upgrade to a larger PV system providing further generation that can utilise a larger capacity battery. Since excess PV generation was previously disincentivised, option (ii) becomes the more cost-effective option. This drives an increase of PV capacity with the installation of battery systems (Fig. 4). Furthermore, low FiT values mean that the loss of FiT revenue (by exceeding the 5 $kW_P$ FiT eligibility limit) does not significantly disincentivise households from installing PV systems > 5 $kW_P$. This results in an earlier transition to PV-battery systems, with continued increases in generation and storage capacity even beyond the high-FiT scenarios. The net-effect with low-FiTs are that households remain economically driven to install PV-battery system capacities that maximise their value from self-consumption as system costs decline.

**In high-FiT scenarios** ($FiT_{75}$ and $FiT_{100}$), the FiT value is greater than the value of self-supply (i.e., $\$_{FiT} > \$_{SS}$). With FiTs being more valuable, exceeding the 5 $kW_P$ FiT eligibility limit becomes a stronger disincentive, as losing all FiT revenue becomes more financially significant. Energy storage has limited financial advantage, as it would swap higher valued excess generation ($\$_{FiT}$) for lower valued self-consumption ($\$_{SS}$). The higher value of grid-exports means that low consumption households can cost-effectively invest in a larger PV system (beyond their self-consumption needs) up to 5 $kW_P$ eligibility limit. Conversely, high consumption households are disincentivised from installing systems larger than 5 $kW_P$ (or they would lose FiT revenue) while also receiving proportionally less revenue from excess generation. These conditions cause the average installed PV capacity per household to rapidly converge towards 5 $kW_P$. As retail tariffs increase and system costs decrease, an increasing number of high consumption households (that have higher electricity bills) eventually find it more cost-effective to reduce their electricity bills by increasing self-consumption (with additional PV capacity and large capacity batteries), over artificially limiting their excess generation with a 5 $kW_P$ system. A higher cost of energy and lower system cost is needed to breakeven, thus delaying the transition towards PV-battery systems (Fig. 4). The net-effect with high FiTs, is that households are initially driven to maximise grid-exports up to the 5 $kW_P$ FiT eligibility limit, but eventually (as system costs decline) a growing percentile of high consumption households find it more cost-effective to forego FiT revenue and maximise their value from self-consumption as system costs decline.

**In the $FiT_{50}$ scenario,** the value of self-supply is equal to the FiT (i.e., $\$_{SS} = \$_{FiT}$) and the resulting investment dynamics comprise of a mix of the high- and low-FiT scenarios (Fig. 4). The first decade of the simulation mirrors the high-FiT scenarios, where the FiT and eligibility limit is sufficient to incentivise the majority of households to invest in 5 $kW_P$ PV systems, while the final decade of the simulation generally mirrors the low-FiT scenarios with a larger amount of pre-installed PV capacity. A transition period between 2030 and 2034 occurs where more households that invest in battery systems choose to retain (rather than increase) their existing PV capacity, leading to a temporary reduction of annual grid-exports. However, from 2035 onwards an increasing majority of households find it more cost-effective to forego FiT revenue with larger PV and battery systems that maximise their value from self-supply over grid-exports.

Across all five FiT scenarios, the flat tariff structure with decreasing PV battery costs and increasing retail electricity tariffs, eventually incentivises households to invest in PV-battery systems (Fig. 4) that maximise the value of self-consumption over excess generation. This leads to households foregoing FiT revenue and gradually investing in PV capacities above 5 $kW_P$ with associated battery storage. This common outcome aligns each FiT scenarios' transition pathways into a set of corresponding grid-operation stages that are used in Section 5 to qualitatively assess the impact of household PV-battery investments on the wider power sector. Two additional sensitivity cases, higher and lower growth retail conditions (Appendix B), were evaluated to assess the robustness of this outcome. The overall qualitative patterns were maintained but with a slower rate of transition in the low growth case and a faster rate of transition in the high growth case.

### 4.1. Emergence of grid-operation stages

Each FiT scenario leads to a different amount of installed PV and battery capacity within each year of the simulation (Fig. 4), but all scenarios lead to a transitional tipping point from household PV-only to PV-battery adoption. As underlying demand for each household is assumed to remain consistent each year, grid-utilisation changes are therefore driven by installed PV and battery capacities. Using the average installed PV and battery capacity per household as independent classifiers (Table 2), we characterise changes in grid-utilisation into a series of distinct grid-operation stages (Table 3). For example, if the average installed PV capacity is 3.5 $kW_P$ (PV-Small) and the average installed battery capacity is 14.5 kWh (Battery-Medium) per household, the resulting grid-operation stage is categorised as '$PV_S:B_M$'. If the average installed PV capacity increases to 5 $kW_P$ (PV-Medium), then the subsequent grid-operation stage becomes '$PV_M:B_M$'. Changes between these grid-operation stages establishes a broader transition pathway from PV-only to PV-battery households (Fig. 6). As the electricity system and its market operates higher resolution timescales, these grid-operation stages also provide a set of high resolution (30-min timestep) grid-utilisation profiles that establish changes in diurnal and seasonal grid demand. The combination of the broader transition pathway with the diurnal and seasonal changes, provides the numerical foundation for a qualitative discussion on its wider power sector effects (Section 5).

**Table 2**
Classification ranges of average PV and battery system capacities per household.

| Range label | Capacity range |
|---|---|
| *Average installed PV capacity per household ($kW_P$)* | |
| PV-small ($PV_S$) | 0.5 – 4 |
| PV-medium ($PV_M$) | 4 – 8 |
| PV-large ($PV_L$) | 8 – 12 |
| *Average installed battery capacity per household (kWh)* | |
| Battery-small ($B_S$) | 0.5 – 10 |
| Battery-medium ($B_M$) | 10 – 20 |
| Battery-large ($B_L$) | 20 – 30 |

*Publishing notes:*
*Display Table 2 over a single column*
*Table provided in "Table_portrait.docx"*

In total, seven grid-operation stages are identified (Table 3) that range from two PV-only stages ($PV_S$ and $PV_M$) through to five PV-battery stages ($PV_S$:$B_S$, $PV_M$:$B_S$, $PV_M$:$B_M$, $PV_M$:$B_L$ and $PV_L$:$B_L$). As each grid-operation stage corresponds to a range of PV and battery capacities, specific scenario-years are selected in Table 3 as representatives for each grid-operation stage. The grid-operation stages are evaluated and illustrated individually in Appendix A to determine their diurnal and seasonal operational characteristics. These quantitative changes in grid-operation at the diurnal and seasonal scales are summarised in Table 4. The results show that ongoing investments by households in PV battery systems can significantly change grid-utilisation across a range of operational dimensions that affect the electricity system and its wholesale market.

**Table 3**
Grid-operation stages (based on the average PV and battery system capacities per household) for each FiT scenario. Starred scenario-years (*) are used as the representative for each grid-operation stage.

| Year | $FiT_0$ | $FiT_{25}$ | $FiT_{50}$ | $FiT_{75}$ | $FiT_{100}$ |
|---|---|---|---|---|---|
| 2018 | $PV_S$ | $PV_S$* | $PV_S$ | $PV_M$ | $PV_M$ |
| 2019 | $PV_S$ | $PV_S$ | $PV_M$ | $PV_M$ | $PV_M$ |
| 2020 | $PV_S$ | $PV_S$ | $PV_M$ | $PV_M$ | $PV_M$* |
| 2021 | $PV_S$ | $PV_S$ | $PV_M$ | $PV_M$ | $PV_M$ |
| 2022 | $PV_S$ | $PV_S$ | $PV_M$ | $PV_M$ | $PV_M$ |
| 2023 | $PV_S$ | $PV_S$ | $PV_M$ | $PV_M$ | $PV_M$ |
| 2024 | $PV_S$:$B_S$ | $PV_S$:$B_S$ | $PV_M$ | $PV_M$ | $PV_M$ |
| 2025 | $PV_S$:$B_S$ | $PV_S$:$B_S$* | $PV_M$ | $PV_M$ | $PV_M$ |
| 2026 | $PV_S$:$B_S$ | $PV_M$:$B_S$ | $PV_M$ | $PV_M$ | $PV_M$ |
| 2027 | $PV_M$:$B_S$ | $PV_M$:$B_S$* | $PV_M$ | $PV_M$ | $PV_M$ |
| 2028 | $PV_M$:$B_S$ | $PV_M$:$B_S$ | $PV_M$ | $PV_M$ | $PV_M$ |
| 2029 | $PV_M$:$B_M$ | $PV_M$:$B_S$ | $PV_M$ | $PV_M$ | $PV_M$ |
| 2030 | $PV_M$:$B_M$ | $PV_M$:$B_M$* | $PV_M$:$B_S$ | $PV_M$ | $PV_M$ |
| 2031 | $PV_M$:$B_M$ | $PV_M$:$B_M$ | $PV_M$:$B_S$ | $PV_M$ | $PV_M$ |
| 2032 | $PV_M$:$B_M$ | $PV_M$:$B_M$ | $PV_M$:$B_S$ | $PV_M$:$B_S$ | $PV_M$ |
| 2033 | $PV_M$:$B_L$ | $PV_M$:$B_M$ | $PV_M$:$B_M$ | $PV_M$:$B_S$ | $PV_M$:$B_S$ |
| 2034 | $PV_M$:$B_L$ | $PV_M$:$B_L$ | $PV_M$:$B_M$ | $PV_M$:$B_S$ | $PV_M$:$B_S$ |
| 2035 | $PV_M$:$B_L$ | $PV_M$:$B_L$* | $PV_M$:$B_M$ | $PV_M$:$B_S$ | $PV_M$:$B_S$ |
| 2036 | $PV_L$:$B_L$ | $PV_M$:$B_L$ | $PV_L$:$B_L$ | $PV_M$:$B_S$ | $PV_M$:$B_S$ |
| 2037 | $PV_L$:$B_L$ | $PV_L$:$B_L$* | $PV_L$:$B_L$ | $PV_M$:$B_S$ | $PV_M$:$B_S$ |

*Publishing notes:*
*Display over a single column (preferable)*
*Table provided in "Table_portrait.docx"*
*Are coloured tables permitted? If not, please revert to black and white*

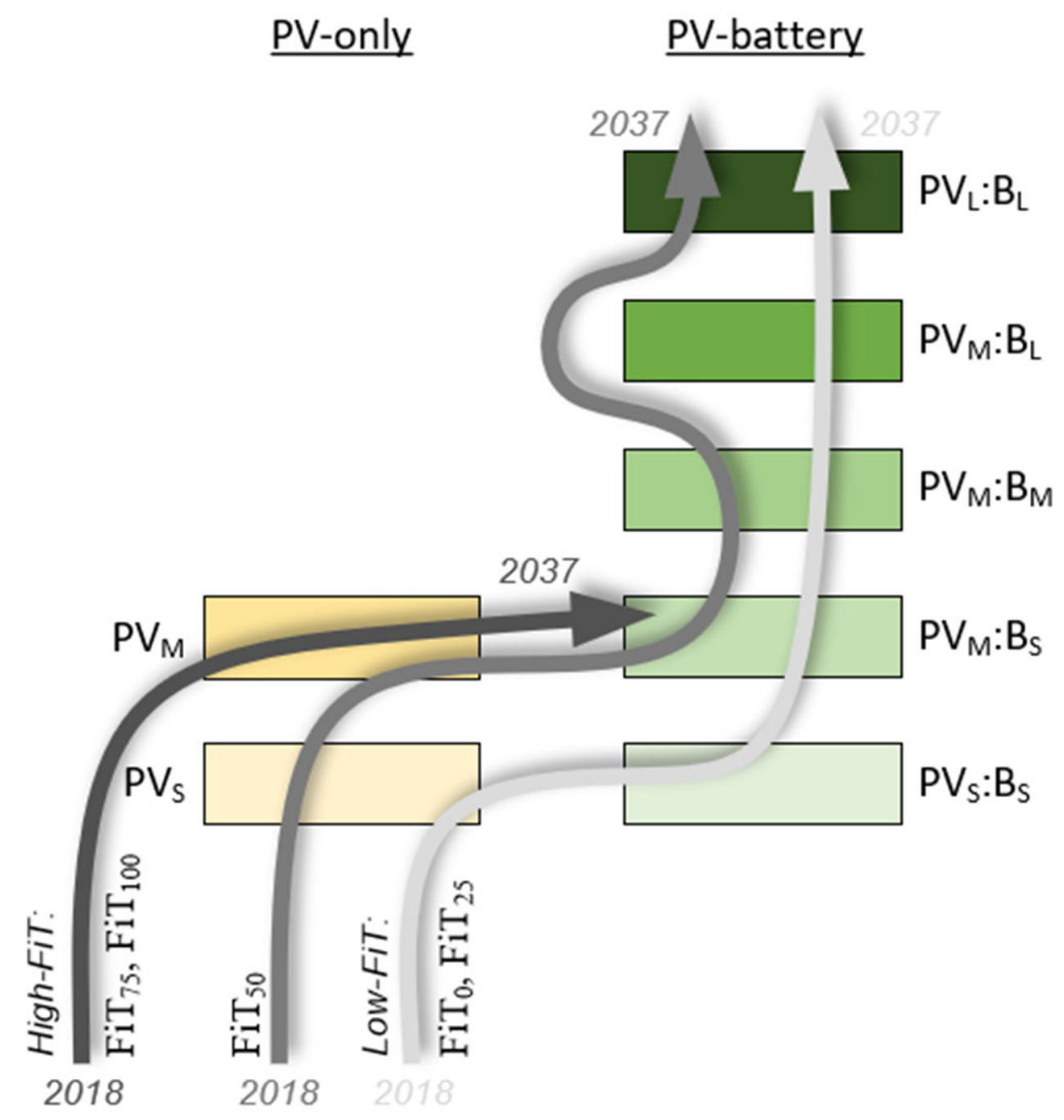

**Fig. 6.** Transition paths (2018 to 2037) for each FiT scenario across the various grid-operation stages.

<u>Publishing notes:</u>
Display over a single column.
Vector graphic provided in
" Figure 6.eps"

**Table 4**
Summarised operational results from each representative grid-operation stage.

| Description | Underlying | $PV_S$ | $PV_M$ | $PV_S$:$B_S$ | $PV_M$:$B_S$ | $PV_M$:$B_M$ | $PV_M$:$B_L$ | $PV_L$:$B_L$ |
|---|---|---|---|---|---|---|---|---|
| Representative scenario *(Year)* | n/a | $FiT_{25}$ (2018) | $FiT_{100}$ (2020) | $FiT_{25}$ (2025) | $FiT_{25}$ (2027) | $FiT_{25}$ (2030) | $FiT_{25}$ (2035) | $FiT_{25}$ (2037) |
| Average PV capacity per household ($kW_P$) | 0 | 1.23 | 4.98 | 3.64 | 4.97 | 5.96 | 7.59 | 8.29 |
| Average battery capacity per household (kWh) | 0 | 0 | 0 | 3.57 | 5.94 | 12.16 | 22.33 | 24.94 |
| Annual grid-imports (MWh) *(% grid dependency)* | 1466 *(100%)* | 1086 *(74%)* | 912 *(62%)* | 721 *(49%)* | 559 *(38%)* | 340 *(23%)* | 174 *(12%)* | 151 *(10%)* |
| Annual peak demand (kW) *(Season)* | 663 *(Summer)* | 654 *(Summer)* | 643 *(Summer)* | 565 *(Summer)* | 519 *(Summer)* | 400 *(Winter)* | 364 *(Winter)* | 349 *(Winter)* |
| Timing of diurnal peak demand *(% occurrence)* | Late afternoon:<br>17:30 – 21:00<br>*(97%)* | Late afternoon:<br>17:30 – 21:00<br>*(97%)* | Late afternoon:<br>17:30 – 21:00<br>*(97%)* | Evening:<br>*Main peak*<br>20:00 – 21:30<br>*(35%)*<br>*distributed over*<br>18:30 – 23:00<br>*(63%)*<br>+<br>Early morning:<br>06:00 – 07:30<br>*(26%)* | Early morning:<br>05:30 – 07:30<br>*(53%)*<br>+<br>Evening/night:<br>20:00 – 23:00<br>*(31%)* | Early morning:<br>05:30 – 08:00<br>*(64%)*<br>+<br>Evening/night:<br>20:00 – 23:00<br>*(24%)* | Early morning:<br>05:30 – 07:30<br>*(40%)*<br>+<br>Evening/night:<br>19:30 – 00:30<br>*(37%)* | Evening/night:<br>19:30 – 01:00<br>*(43%)*<br>+<br>Early morning:<br>05:30 – 07:30<br>*(35%)* |
| Annual grid-exports (MWh) | 0 | 75 | 1139 | 480 | 701 | 737 | 1007 | 1166 |
| Annual peak feed-in (kW) *(% of underlying annual peak demand)* | 0 *(0%)* | 145 *(22%)* | 914 *(138%)* | 611 *(92%)* | 865 *(130%)* | 1021 *(154%)* | 1303 *(196%)* | 1438 *(217%)* |
| Timing of diurnal minimum demand *(% occurrence)* | Early morning:<br>02:30 – 05:30<br>*(96%)* | Midday:<br>10:30 – 14:30<br>*(86%)* | Midday:<br>10:30 – 15:00<br>*(94%)* | Midday:<br>11:30 – 15:00<br>*(92%)* | Midday:<br>11:30 – 15:30<br>*(94%)* | Midday:<br>11:30 – 16:00<br>*(96%)* | Afternoon:<br>12:00 – 16:00<br>*(94%)* | Afternoon:<br>12:00 – 15:30<br>*(91%)* |
| Peak ramp up rate (kW/min) | 3.55 | 3.65 | 10.53 | 6.62 | 8.85 | 9.67 | 12.67 | 14.49 |
| Peak ramp down rate (kW/min) | -3.43 | -3.43 | -9.62 | -4.87 | -6.52 | -6.49 | -7.28 | -8.17 |

*Publishing notes:*
*Displayed over full width, or landscape if it is still too large for the page*
*Table provided in "Table_landscape.docx"*

# 5. Discussion

By characterising the grid-operation stages, a pathway of transition (Fig. 6) is identified that leads a range of operation effects (Table 4) at the aggregate household level. From the use of real underlying demand and PV generation data from 261 households, the aggregate changes in grid-utilisation are assumed to be representative of PV and battery investing households in the SWIS network. As the paper focuses on transition pathway patterns (rather than forecasts), we use the trajectory of operational changes (Table 4) to qualitatively assess the how layers of the power sector are affected starting from an aggregate household perspective through to the system and market layers.

## 5.1. Aggregate household perspective

### *5.1.1. Rising annual peak feed-in with household PV-battery adoption*

During the period of PV-only adoption, both the $PV_S$ and $PV_M$ stages (Appendix A.2 and A.3 respectively) exhibit demand-side changes consistent with the "duck curve" (Denholm et al., 2015; Maticka, 2019). At the diurnal scale, minimum demand shifts from night into midday and becomes increasingly negative (Fig. A.2d and Fig. A.3d). At the seasonal scale, net-exports increase in magnitude over the summer months (Fig. A.2a and Fig. A.3a). As households' transition from PV-only to PV-battery systems, further changes become evident. Notably, from $PV_M$:$B_S$ onwards, the average PV capacity per household increases past the 5 $kW_P$ FiT eligibility limit (Table 4), meaning that grid-exports no longer have financial value (increasing the value of self-consumption) which then drives further PV and battery capacity growth. Considering the gradual changes in grid-utilisation from the $PV_S$:$B_S$ (Fig. A.4a) through to the $PV_L$:$B_L$ (Fig A.8a) stage, there is a gradual reduction in the amount of grid consumption over winter, which indirectly causes grid-exports to rise significantly across the summer months. The use of flat tariff structures leads to a series of economically rational decisions. Firstly, underlying energy demand in winter is higher than summer due to more consistent occurrences of night-time heating demand (Fig. A.1a). Secondly, reduced solar resources and increased night-time demand over winter requires larger PV capacities to raise self-generation during these months. Thirdly, installing battery capacities larger than would be regularly utilised over the entire year leads to diminishing returns that disincentivise households from installing larger storage capacities. During the summer months, this leads to many household batteries becoming full before midday and allowing peak

PV generation to continue feeding into the grid at noon. With PV capacities rising to cover a growing portion of winter demand, peak feed-in during the summer months eventually exceeds the underlying annual peak demand of 663 kW from the $PV_M$:$B_S$ stage onwards (Table 4).

As flat tariffs do not provide an incentive to change the timing of grid-exports, peak feed-in across the majority of households temporally coincide around noon. As the capacity of the distribution network is designed around the expected annual peak demand plus a reserve margin, increasing peak feed-in from further PV-battery investments exacerbates existing hosting capacity limitations and can lead to reverse power flows beyond the capacity of the distribution network. As system and network operators do not currently have the ability to control behind-the-meter generation, risk mitigation and management strategies would have to be taken, such as restricting grid exports, further network augmentation, installing distribution-scale energy storage, or providing dynamic export limits.

#### *5.1.2. Emergence of an early-morning diurnal peak demand*

As is typical of Australian households (AEMO, 2018), the underlying diurnal peak demand occurs most frequently during the late-afternoon between 17:30 and 21:00 (Table 4 and Fig. A.1d). Household PV-only and PV-battery systems affect the timing and magnitude of the diurnal peak demand. In the PV-only stages ($PV_S$ and $PV_M$) the setting sun limits the ability of PV generation to reduce diurnal peak demand, and the late-afternoon peak can only be delayed and reduced slightly (Fig. A.2d and Fig. A.3d). In the PV-battery stages the timing of diurnal peak demand becomes much more sensitive to variations in insolation and installed PV battery capacities. Starting from the $PV_S$:$B_S$ stage, the lower generation and storage capacity means that on days with less than ideal insolation, the energy self-generated and stored only delays the late-afternoon peak to around 20:00 (Fig. A.4d). But on days with higher insolation, there is sufficient energy self-generated and stored that households are able to self-supply past the underlying late-afternoon peak and into the night, thus temporarily eliminating the late-afternoon diurnal peak demand in the process. Once battery storage capacity is exhausted however, grid-imports are required overnight and into the next morning. These factors lead to the first occurrences of an early-morning diurnal peak demand (Table 4). As PV-battery investments progress from the $PV_M$:$B_S$ to the $PV_M$:$B_L$ stage, battery storage capacity increases more than PV capacity. Therefore, on more days of the year, households are able to self-supply further into the night, resulting in an increasing occurrence of the early-morning diurnal peak demand (Fig. A.5d, Fig. A.6d and Fig. A.7d). By the $PV_L$:$B_L$ grid-operation stage however, the much larger self-generation and storage capacity allows households to increasingly self-supply through the night and into the next morning, which then removes grid demand over a diurnal cycle and begins to reduce the occurrence of the early-morning diurnal peak demand (Fig. A.8d).

#### *5.1.3. Shifting into winter dominant residual demand*

Compared to the winter months, the higher levels of summer insolation increase the capability of households to self-supply. Up until the $PV_M$:$B_S$ stage the annual peak demand remains in summer (Table 4). With increasing household PV battery capacities however, the residual demand profile becomes increasingly winter dominant (e.g. Fig. 6a). From $PV_M$:$B_M$ onwards, the summer peak is reduced sufficiently that the annual maximum is replaced by the winter peak (Table 4). Considering monthly grid consumption (Fig. 7), the high insolation levels over summer (Dec to Feb) allow households to reduce a significant portion of their overall grid-imports. In the autumn (Mar to May) and spring (Sep to Oct) months, milder weather conditions reduce heating and cooling demand and when coupled with moderate insolation levels, households are able to reduce their grid-imports beyond the summer months. Low insolation levels in the winter months (Jun to Aug) prevents PV-battery systems from operating as effectively, hence grid-imports remain highest over the winter period.

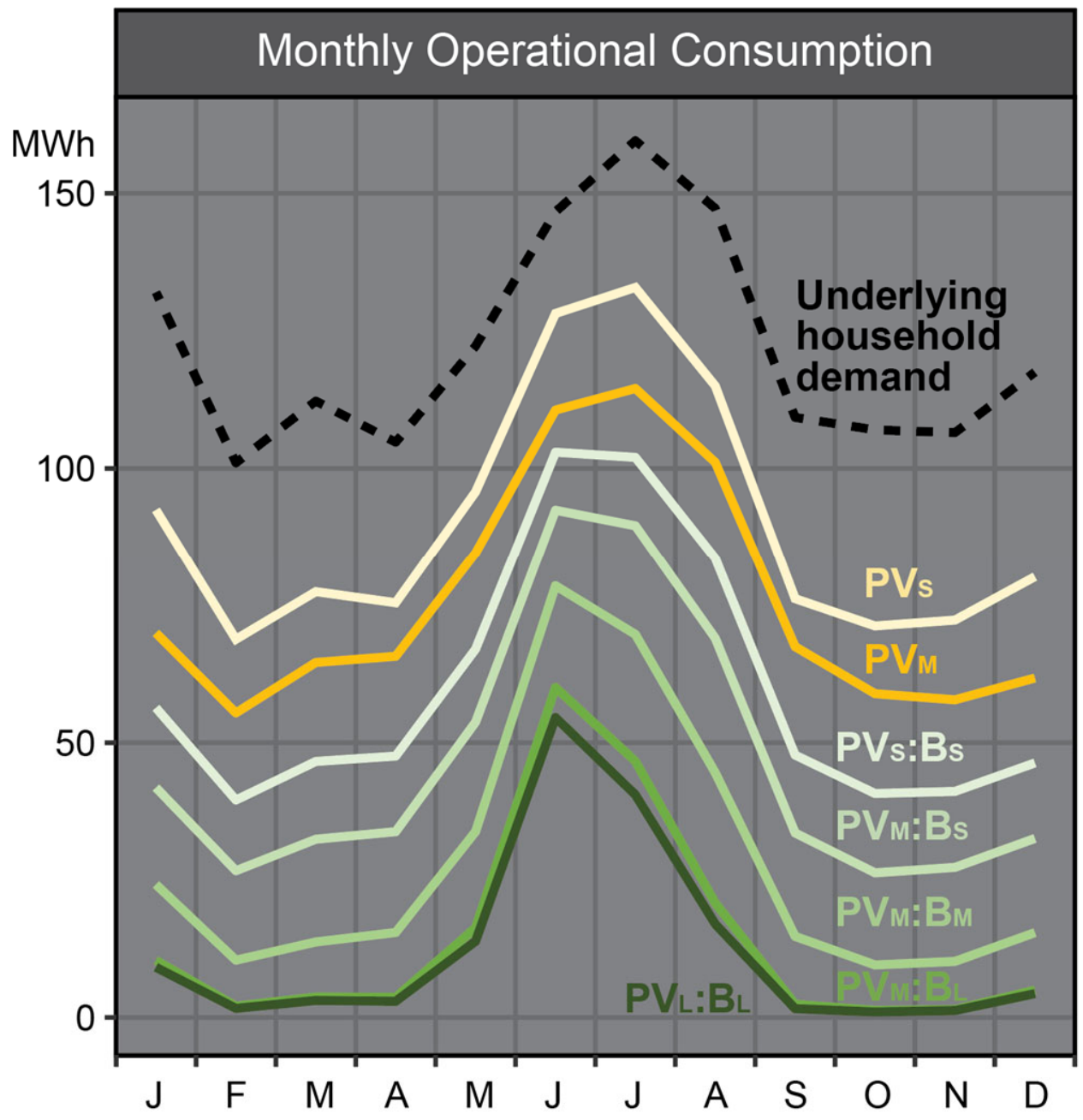

**Fig. 7.** Monthly grid-imports (aggregate of 261 households) of the underlying household and residual demand from each grid-operation stage.

Publishing notes:
Display over a single column.
Vector graphic provided in " Figure 7.eps"

## 5.2. System and market perspectives

Even though households are only one segment of customers contributing to the total grid demand, growing household investments into PV battery systems are still capable of significantly reshaping how the electricity system and market operates. Using the qualitative impacts from the aggregate household analyses in Section 5.1 and contextualising it as a proportion of customers within the total grid demand, we further analyse how the operational and market layers are affected as households transition from PV-only to PV-battery systems under flat retail tariffs (Fig. 6).

### *5.2.1. Operational challenges and opportunities*

#### *5.2.1.1. PV-battery households becoming net-generators*

PV-only systems must continue to rely on grid-sourced energy at night, ensuring a minimum level of grid demand is always maintained. PV-battery systems however can continue to self-supply much further into the night, leading to additional reductions in annual grid-imports (Fig. 4). Furthermore, annual grid exports continue to rise with household PV-battery adoption (Table 4). The net effect is that annual grid-exports from PV-battery households eventually exceed annual grid-imports, making these households become net-generators and a growing source of renewable energy generation. Such an outcome, if widespread, fundamentally challenges traditional liberalised electricity markets, with PV-battery households competing with and displacing utility generation while avoiding fixed electricity system costs. Continued reductions in annual grid-imports also places downward pressure on any growth of total grid demand, which disincentivises future investments into additional bulk-energy utility generation.

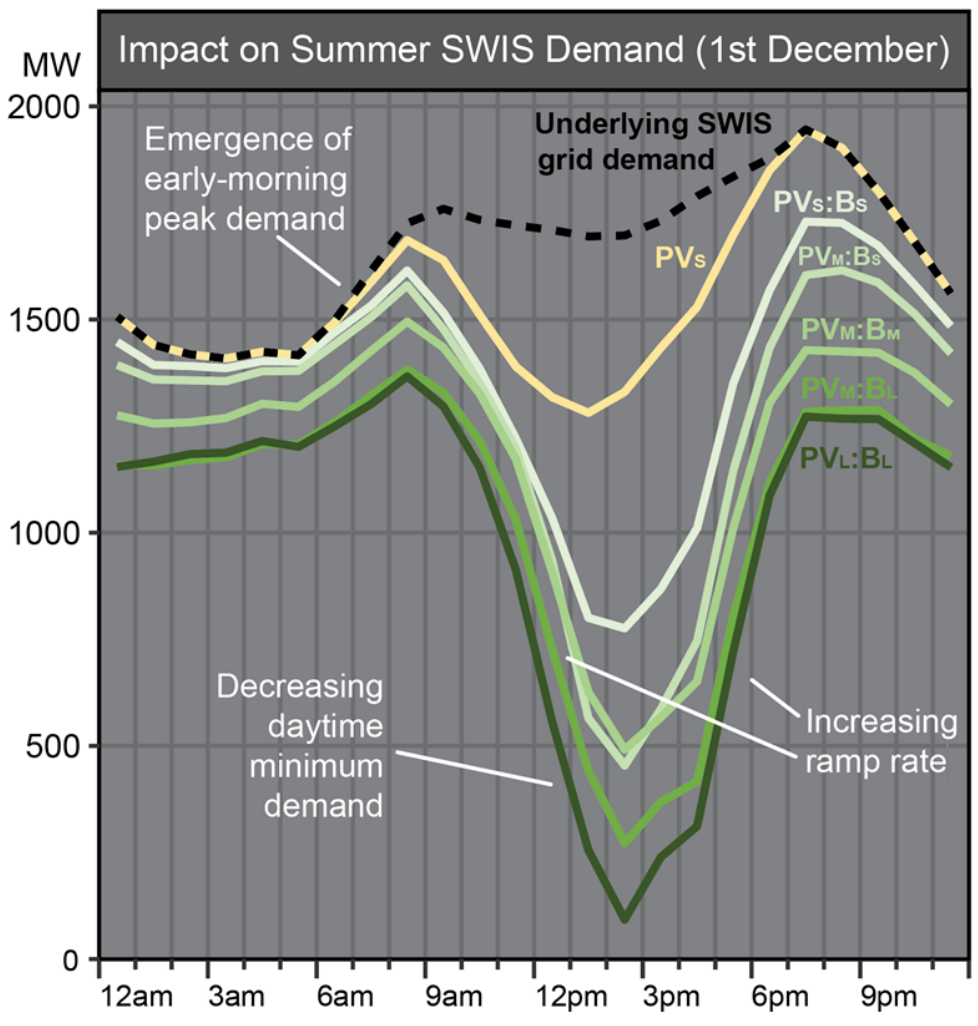

**Fig. 8.** Impact on summer diurnal demand (1st December) from 270,000 PV-only households transitioning to PV-battery households in the $FiT_{25}$ scenario and using real SWIS network demand (AEMO, 2019c).

Publishing notes:
Display over a single column.
Vector graphic provided in " Figure 8.eps"

#### 5.2.1.2. Changes in timing and ramping of utility-scale generation

As household PV-battery investments progress through each grid-operation stage (Fig. 6), daytime grid feed-in continues to rise (Section 5.1.1), the late-afternoon diurnal peak demand subsides and is gradually replaced by an early-morning diurnal peak (Section 5.1.2). A representation of these residual demand changes on the total grid demand in summer[16] is shown in Fig. 8. Using a rooftop PV penetration rate of 27% (of 1 million households) in the SWIS network (APVI, 2019a), changes in residual demand become clearly capable of affecting the total grid demand:

(i) Overall minimum demand continues to decline as daytime feed-in from PV-battery households increase (with household batteries unable to store all excess PV generation);

(ii) The reduction and shift of diurnal peak demand from the late-afternoon to early morning, by PV-battery households, become capable of affecting the whole system and driving similar changes in the diurnal profile of the total grid demand.

Even though overall diurnal peak demand is reducing, the minimum demand during noon continues to decline, leading to ramp rate increases (Fig. 8). This suggests that as household PV-battery systems become more widespread, the appropriate types of utility generation may be affected. With decreasing annual grid demand (Section 5.2.1) and reductions in diurnal peak demand, there is a reduced need for additional utility generation capacity. However, with increasing ramp rates, flexible generators have an increasing advantage over inflexible baseload generators.

---

[16] The diurnal SWIS grid profile was obtained via SCADA records (AEMO, 2019c) on 1st December 2012, as it best matches the household data profiles (Ausgrid, 2018) collected between 1st July 2012 and 30th June 2013.

#### 5.2.1.3. Coordination increasingly necessary for household PV-battery assets

These changes in total grid demand are influenced by the continued use of flat retail tariffs. Without a temporal value, there is no financial benefit for households to operate their DER systems according to the dynamic needs of the electricity system. Currently household DER systems are not centrally monitored or controlled (AEMO, 2019a); hence system and market operations are reactive to household PV-battery adoption and dispatch. Addressing system limitations at the utility-side (e.g. additional peaking generation, utility-scale energy storage, network augmentation) may be more costly than managing household energy resources directly. By developing the capability for system coordination behind-the-meter, not only are operational risks and mitigation costs reduced but a pathway for customers to provide a wider range of energy services becomes available. Furthermore, as household PV battery capacities increase, underutilised generation and storage capacity becomes available behind-the-meter to supply and manage a growing share of overall electricity demand. These changes are likely to involve significant regulation, privacy, and market reforms (AEMC, 2019; AEMO and Energy Networks Australia, 2018) before these behind-the-meter services can be integrated into the electricity network and market.

### 5.2.2. Electricity market challenges and opportunities

#### 5.2.2.1. Falling retailer revenues necessitates DER market integration

As PV-battery systems can achieve much greater grid-import reductions than PV-only systems,[17] retailers that collect revenues primarily from volumetric usage charges are exposed to significant lost sales from widespread household PV-battery adoption. The transition from PV-only to PV-battery systems also leads to many households foregoing FiT revenues and investing in PV-battery systems with PV capacities above the 5 kWP FiT eligibility limit (Section 4). As retailers no longer have to pay for household grid-exports, and household PV-battery systems are already sunk costs, any grid services that could be provided by these DER assets (e.g., peak shaving, frequency response, load shifting) has near-zero marginal costs. This creates an opportunity for retailers to reposition themselves from managing the risk of wholesale electricity prices to becoming an agent for DER market participation and encouraging their integration into the wholesale electricity market. However, considerations have to be made within electricity market rules to allow household DER systems to effectively compete for grid services and to allow their potential system and operational savings to be realised across the system (AEMC, 2019).

[17] In the $PV_M$ stage only 38% of grid consumption could be self-supplied, however in the $PV_L$:$B_L$ stage, over 90% of household grid consumption could be self-supplied (Table 4).

##### *5.2.2.2. Increasing role for flexible demand*

Changes in residual demand from household PV-battery systems (Fig. 8) leads to increased ramping of network demand (Section 5.2.1.2) while reducing overall grid consumption (Section 5.2.1.1). Flexible generation technologies that can respond rapidly to these changes in demand, such as peaking and load balancing facilities, should gain a competitive advantage over inflexible baseload generators. But as their levelised costs of electricity are typically higher than baseload generators (Bloomberg New Energy Finance, BNEF, 2019a; Graham et al., 2018), the increased use of flexible generators may place upward pressure on wholesale electricity prices. However, the integration of household DER systems into the electricity market as a form of flexible demand (with near-zero marginal costs) may provide a competitive alternative to utility-scale peaking and load balancing facilities.

## 5.3. Limitations and outlook

The results remain dependent on the choice of input parameters and modelling assumptions. A business-as-usual perspective was taken to assess the range of impacts that may emerge from households PV battery investments without endogenous feedback on the potential changes in the wholesale energy market and associated retailer and network costs. This leads to a range of limitations to be considered when interpreting the results.

*Underlying electricity demand and solar insolation profiles are repeated year-on-year.* By keeping these parameters constant, the results can more clearly show the influence of the retail conditions on household PV battery adoption. However, it also discounts future changes in energy demand, energy efficiency, climatic conditions and ignores the potential of electric vehicles to further reshape the underlying electricity demand. As the demand and solar insolation profiles reflect the specificities of the region, caution is required when transposing results into other regions with different demand and insolation profiles.

*Flat usage charge and feed-in tariffs and the default battery operation.* Australian retail electricity tariffs are predominantly time-invariant, consisting of a usage and fixed charge (without demand charges). Without a temporal value of energy, batteries are not incentivised to operate beyond improving self-consumption (i.e., default battery operation[18]), as additional layers of operational complexity (e.g., deciding when to grid-charge and grid-discharge) would incur additional costs without further remuneration. The use of time-varying tariffs (that would encourage different operational behaviour) remains outside the scope of this paper.

[18] The default mode of battery operation maximises PV self-consumption by, only charging using excess PV generation until full, and only discharging to avoid grid-imports until empty (while remaining within the 5 kW battery inverter limit). As a result, grid-charging and grid-discharging operation is not utilised.

*Uniform household PV battery investment methodology.* Results were generated by applying a single investment methodology (based on discounted cash flows from bill savings) to each household. The decision process reflects economically rational homeowners with sufficient income that can finance PV battery investments by extending their existing home loans. With a uniform investment methodology, these results cannot reflect the full spectrum of factors influencing customer PV battery adoption, nor the wide range of financial valuation metrics that households may use to make their investment decisions. However, empirical evidence continues to reaffirm that bill savings (Agnew and Dargusch, 2017; Bondio et al., 2018; Figgener et al., 2019) and homeownership (Sommerfeld et al., 2017) are significant factors that influence the installation of behind-the-meter energy systems, and discounted cash flows remain widely used in the literature (e.g., Schram, 2018; Schopfer; 2018).

*Focus on households.* As households remain the largest customer sector installing behind-the-meter PV systems (AEMO, 2019d), this paper focuses on the retail market conditions that affect their PV battery investments. With ownership being a significant influencing factor (Sommerfeld et al., 2017), commercial and industrial (C&I) customers (that are largely tenanted) are disadvantaged from installing DER assets, since risk and benefit sharing between landlords and tenants need to be first established. However, if electricity prices continue to increase, the C&I sector may encourage additional DER growth to reduce their exposure to future price increases and drive another set of electricity system transition patterns.

*Future electricity prices and PV battery system costs.* Projected electricity prices and PV battery system costs are represented using exogenous scenario parameters and change at a fixed rate each year. With rising (time-invariant) electricity prices and declining system costs, these cost projections only reflect historical Australian business-as-usual conditions and industry price expectations. This paper does not consider further cost dynamics, such as continued changes to the electricity system,[19] introduction of new support policies or further expansion of global PV battery supply chains. Exploring how these dynamics interact and affect subsequent policy decisions may be a promising area for future research.

Access to high resolution and disaggregated household consumption and generation profiles would allow researchers and energy analysts to provide further more extensive analysis to policymakers and system managers, allowing them to be better informed on the expected growth of behind-the-meter PV battery systems and their potential to provide grid services. Further research is required to understand the influence of a broad range of retail tariff structures on household PV battery investment behaviours over time and their system integration impacts. Additional research is required to evaluate and quantify the suitability of various utility-scale generation technologies as households adopt PV battery systems. Future research could also assess the policy costs and carbon abatement potential from household PV battery investments over utility-scale solutions.

## 6. Conclusions and policy implications

With behind-the-meter PV battery systems, households effectively have the highest dispatch priority on the network. By changing their grid consumption and freely exporting energy into the grid, these prosuming households have the ability to reshape grid demand, revenue streams and displace utility-scale generation. As households are capable of reacting to retail electricity costs by investing in additional PV and battery capacity, policymakers have to carefully consider how future households should interact with the grid and its role in the electricity system.

[19] Such as, the wider integration of zero-marginal cost generation that can reduce wholesale electricity prices, or the modernisation of distribution networks that may raise network and operation costs.

With flat retail tariffs the transition from PV-only to PV-battery households leads to a significant reduction in grid-imports and significant rise in grid-exports, until households eventually become net-generators. These net-generator households are able to avoid system costs that may be incurred from the additional operational and market responses necessary to accommodate their grid-utilisation changes. This exacerbates inequality by increasing the cost burden on all other customers and introduces market inefficiencies by displacing lower cost generation. Government policymakers may be able to avoid making changes at low penetration rates, but as PV-battery households become more widespread (as is the potential on the SWIS) their power sector impacts can no longer be ignored, which would require action from both government policymakers and power system regulators.

PV-battery households should eventually be treated like other generators that supply electricity to the grid, with responsibilities to provide firm and reliable power when it is required. Either proscriptive or price-based policies can be used. Households that want to supply electricity to the grid should accept a common grid code that provides system-level visibility, aligns their operational dynamics, and allows remote feed-in management. This would allow critical system operation or market prices to determine when electricity can be exported. Recouping costs associated with the negative externalities from PV-battery households (e.g., extra costs imposed on the distribution network, lost retailer revenues) will also be necessary. Increasing fixed daily charges over volumetric usage charges, potentially limits the incentive for further PV battery adoption, but applies to all customers and is thus regressive. An access fee could be applied when exporting to the grid, which places the cost burden only on prosuming households, but discourages PV exports along with their carbon abatement potential. Another approach is to reduce these negative externalities by changing when PV exports occur. For example, by introducing time-varying FiTs[20] that reflect the temporal value of grid-exports (i.e., significantly reducing the value of midday grid-exports and increasing their value during peak hours). Over the short-term this should discourage further increases in PV capacity while also bringing forward the timing and scale of battery systems. By better aligning grid-exports with peak demand rather than minimum operational demand during midday, there is less pressure on wholesale electricity prices. Finally, using retail aggregators[21] to manage household grid-exports and grid-imports in line with the wholesale electricity market, should increase competition for flexible generation and demand, and improve the market's economic efficiency, further lowering prices for all electricity customers. As our analysis has shown, reducing the value of the FiT accelerates PV-battery adoption which improves the capability of households to provide firm capacity. This provides policymakers an opportunity to provide the price signals that encourage prosuming households to better align their load and generation with the needs of the wider electricity system.

This paper illustrates the potential magnitude of changes and challenges that continuous investments by households into PV-battery systems, under flat retail tariffs, may have on the electricity system. These households are effectively using private capital to invest into the power sector, and as the cost-effective tipping point for PV-battery systems approaches, it becomes increasingly important to develop policy strategies that encourage future household investments to complement the electricity system and allow customers to play a bigger role in decarbonising the power sector.

[20] https://www.esc.vic.gov.au/electricity-and-gas/electricity-and-gas-tariffs-and-benchmarks/minimum-feed-tariff
[21] https://homebatteryscheme.sa.gov.au/join-a-vpp

## Acknowledgements

This work was supported by resources provided by The Pawsey Supercomputing Centre with funding from the Australian Government and the Government of Western Australia. The authors would also like to thank the two anonymous reviewers for their constructive and insightful comments.

## Appendix A. Representative grid-operation stages

For each representative stage, the following sub-sections describe the changes to annual grid-operation at a 30-min resolution. The values are summarised and presented in Table 4 with the transitional implications discussed in Section 5.

### A.1. Underlying aggregate household demand

The annual grid-imports from the 261 household customers (Ausgrid, 2018) is 1466 MWh with an annual peak demand of 663 kW (Fig. A.1a and Fig. A.1b) that occurs in summer due to high cooling loads. The average annual consumption is 5.62 MWh per household with an average solar PV capacity factor of 14.8%. The diurnal peak demand typically occurs in the late-afternoon between 17:30 and 21:00 and the diurnal minimum demand typically occurs in the early-morning between 02:30 and 05:30 (Fig. A.1c and Fig. A.1d).

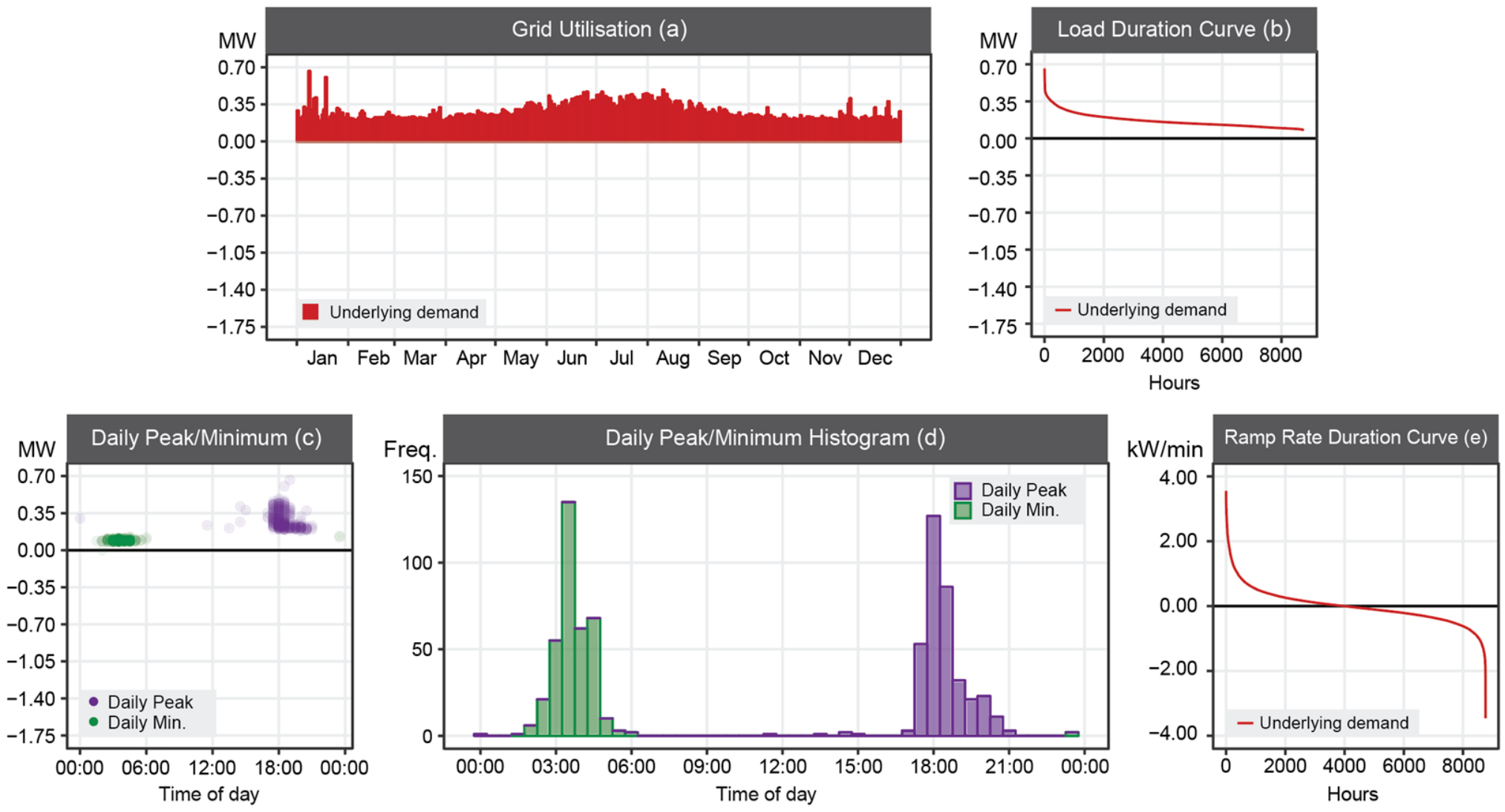

**Fig. A.1.** Annual grid-utilisation (aggregate of 261 households) of the underlying demand. (a) Half-hourly grid-imports. (b) Load duration curve. (c) Capacity and timing of diurnal demand peaks and minimums. (d) Histogram of diurnal demand peaks and minimums across each time interval. (e) Ramp rate duration curve.

Publishing notes:
Display across the whole page.
Vector graphic provided in " Figure A.1.eps"

## A.2. PV-small only ($PV_S$) stage

The $PV_S$ stage is represented with the time-series residual grid-utilisation from the $FiT_{25}$ scenario in the year 2018 (Table 3) with an average PV capacity of 1.23 $kW_P$ and no installed battery capacity (Fig. 4).

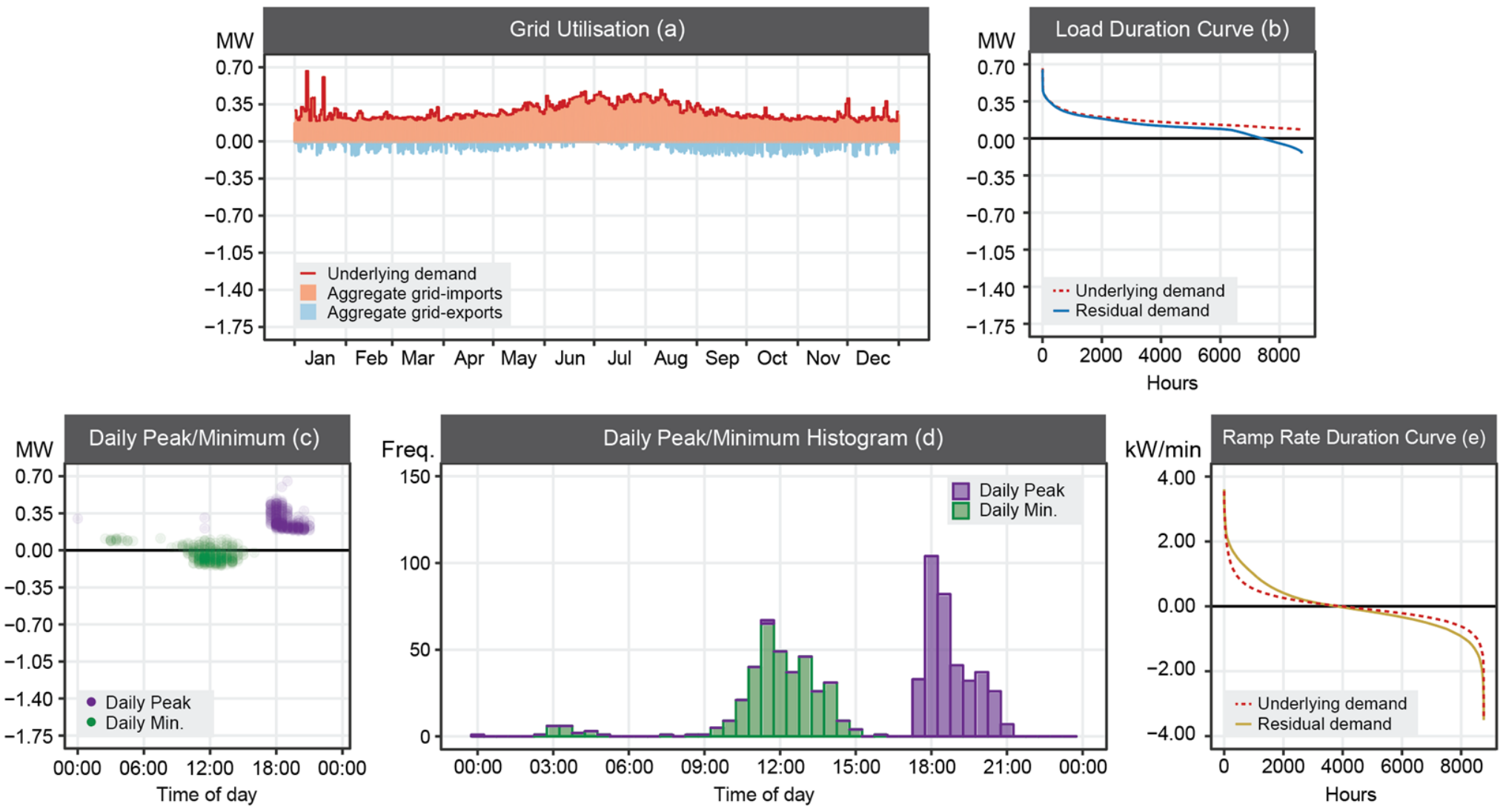

**Fig. A.2.** Annual grid-utilisation (aggregate of 261 households) at the PVs (PV-small) operational stage. (a) Half-hourly grid-imports and grid-exports of residual and underlying demand. (b) Load duration curve of residual and underlying demand. (c) Capacity and timing of diurnal demand peaks and minimums. (d) Histogram of diurnal demand peaks and minimums across each time interval. (e) Ramp rate duration curve of residual and underlying demand.

<u>Publishing notes:</u>
Display across the whole page.
Vector graphic provided in “ Figure A.2.eps”

The grid-utilisation (Fig. A.2a) shows significant grid-imports remaining over the year and annual grid-imports of 1086 MWh. Compared to underlying demand, annual peak demand is slightly reduced to 654 kW (Fig. A.2b) while continuing to occur in summer. Customer annual grid-exports total 75 MWh and peaks at 145 kW (Fig. A.2b). The diurnal peak demand remains predominantly in the late-afternoon between 17:30 at 21:00 (Fig. A.2c) but with 56% more peak demand periods occurring between 19:30 and 21:00 (Fig. A.2d). More significantly, the diurnal minimum demand period moves from a positive value in the early-morning (Fig. A.1c) to an increasingly negative value around midday (Fig. A.2c), the implications of which are discussed in Section 5.1.1.

## A.3. PV-medium only ($PV_M$) stage

The $PV_M$ stage is represented with the time-series residual grid-utilisation from the $FiT_{100}$ scenario in the year 2020 (Table 3) with an average PV capacity of 4.98 $kW_P$ and no installed battery capacity (Fig. 4).

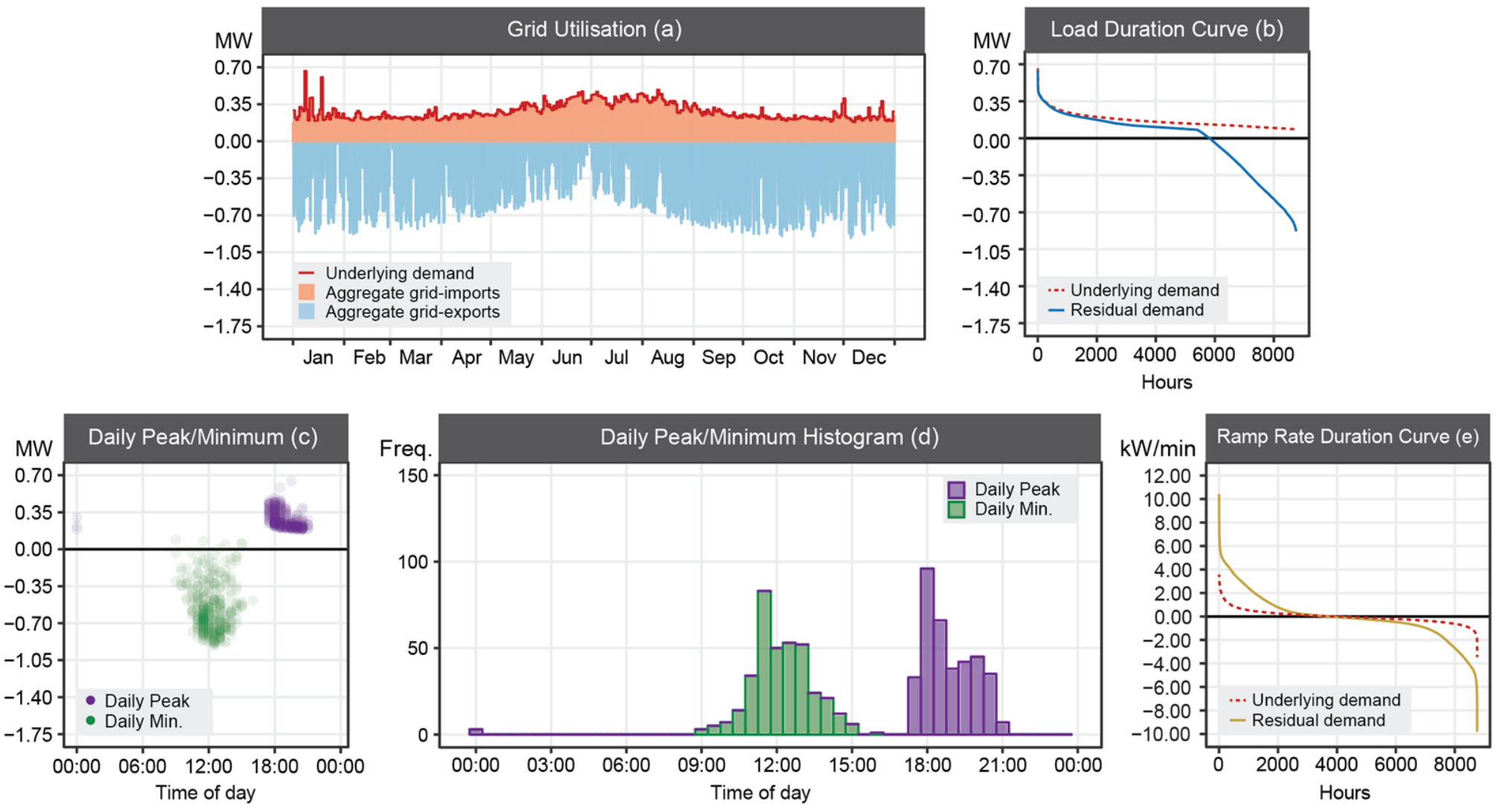

**Fig. A.3.** Annual grid-utilisation (aggregate of 261 households) at the $PV_M$ (PV-medium) operational stage. (a) Half-hourly grid-imports and grid-exports of residual and underlying demand. (b) Load duration curve of residual and underlying demand. (c) Capacity and timing of diurnal demand peaks and minimums. (d) Histogram of diurnal demand peaks and minimums across each time interval. (e) Ramp rate duration curve of residual and underlying demand.

Publishing notes:
Display across the whole page.
Vector graphic provided in " Figure A.3.eps"

Compared to $PV_S$, the greater installed PV capacity in $PV_S$ has further reduced annual grid-imports to 912 MWh (Fig. 4). The grid-utilisation (Fig. A.3a) shows that the majority of grid-imports continues and annual peak demand only slightly reduces to 643 kW (and occurs in summer). The diurnal peak demand period remains predominantly in the late-afternoon between 17:30 at 21:00 (Fig. A.3c) but with a further increase in the evening periods between 19:00 and 21:00 (Fig. A.3d). The additional PV capacity (compared to $PV_S$) significantly increases customer annual grid-exports to 914 MWh (Fig. A.3a) that peaks at 914 kW (Fig. A.3b). This grid-export peak exceeds the underlying peak demand of 663 kW, which has implications for network capacity design (discussed in Section 5.1.1).

## A.4. PV-small and battery-small ($PV_S$:$B_S$) stage

The $PV_S$:$B_S$ stage is represented with the time-series residual grid-utilisation from the $FiT_{25}$ scenario in the year 2025 (Table 3) with an average PV capacity of 3.64 $kW_P$ and an average battery capacity of 3.57 kWh per household (Fig. 4).

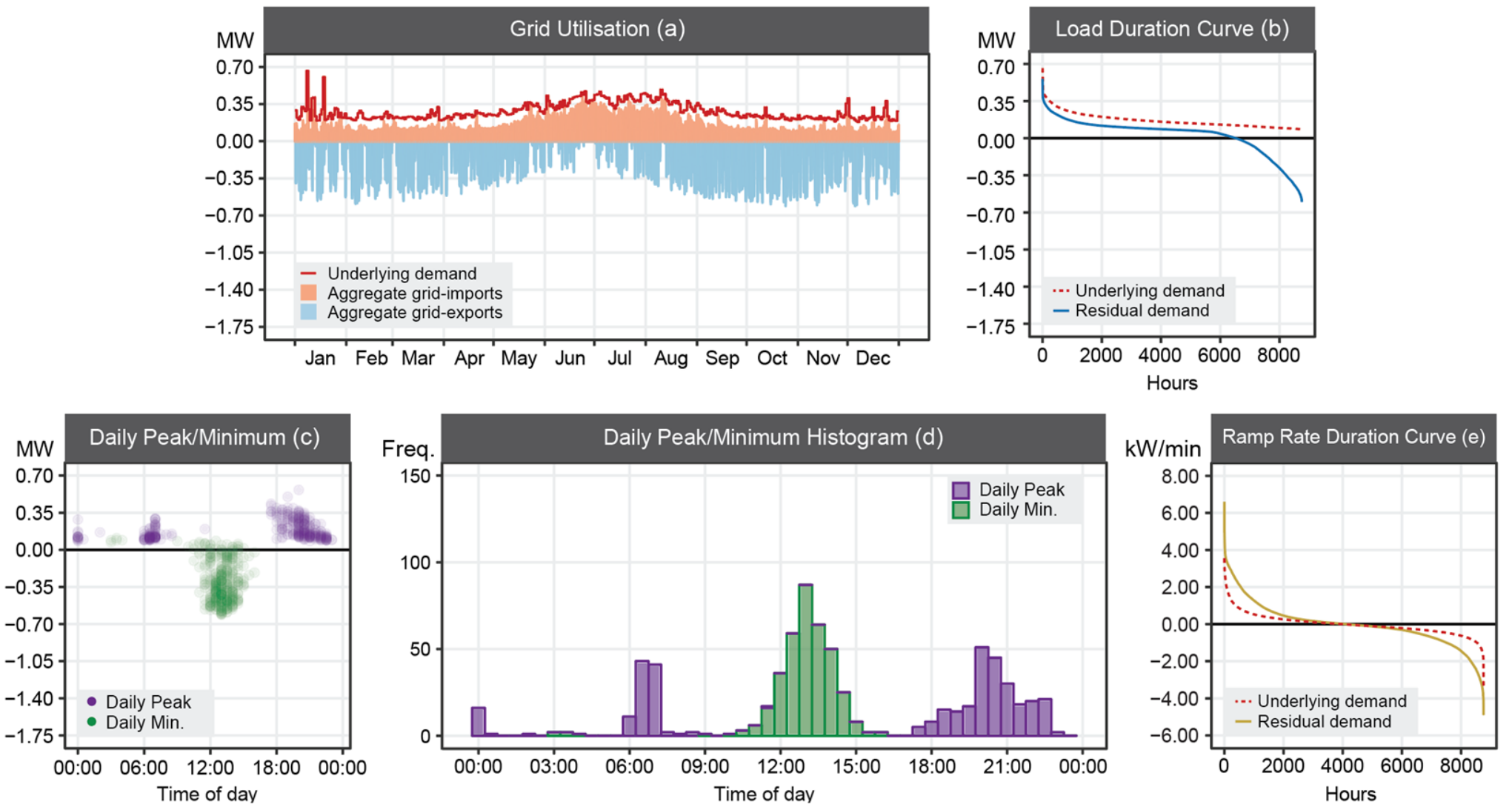

**Fig A.4.** Annual grid-utilisation (aggregate of 261 households) at the $PV_S$:$B_S$ (PV-small and battery-small) operational stage. (a) Half-hourly grid-imports and grid-exports of residual and underlying demand. (b) Load duration curve of residual and underlying demand. (c) Capacity and timing of diurnal demand peaks and minimums. (d) Histogram of diurnal demand peaks and minimums across each time interval. (e) Ramp rate duration curve of residual and underlying demand.

Publishing notes:
Display across the whole page.
Vector graphic provided in " Figure A.4.eps"

Compared to $PV_S$, the widespread adoption of low-capacity battery systems leads to a more consistent reduction in annual grid-imports (Fig. A.4a). Annual grid-imports fall to 721 MWh while annual peak demand reduces more sharply to 565 kW (Fig. A.4b) (remaining in summer). The installation of low-capacity battery systems also raises the average level of installed PV capacity from 1.23 $kW_P$ to 3.64 $kW_P$ per household that increases PV self-generation and improves the overall self-consumption and financial benefits from the PV-battery system (discussed in Section 4). The net effect of the higher installed PV capacity per household is an increase in annual grid-exports to 480 MWh that peaks at 611 kW in summer (Fig. A.4a and Fig. A.4b). The timing of diurnal peak demand shifts from the late-afternoon into two separate time intervals, firstly a wider evening peak (concentrated between 20:00 and 21:30 but distributed over 18:30 and 23:00), and a second early-morning peak (between 06:00 at 07:30). Furthermore, the timing of the diurnal minimum demand is generally delayed by an hour (between 11:30 and 15:00). These changes to diurnal peak and minimum demand are discussed in Section 5.1.2.

### A.5. PV-medium and battery-small ($PV_M$:$B_S$) stage

The $PV_M$:$B_S$ stage is represented with the time-series residual grid-utilisation from the $FiT_{25}$ scenario in the year 2027 (Table 3) with an average PV capacity of 4.97 $kW_P$ and an average battery capacity of 5.94 kWh per household (Fig. 4).

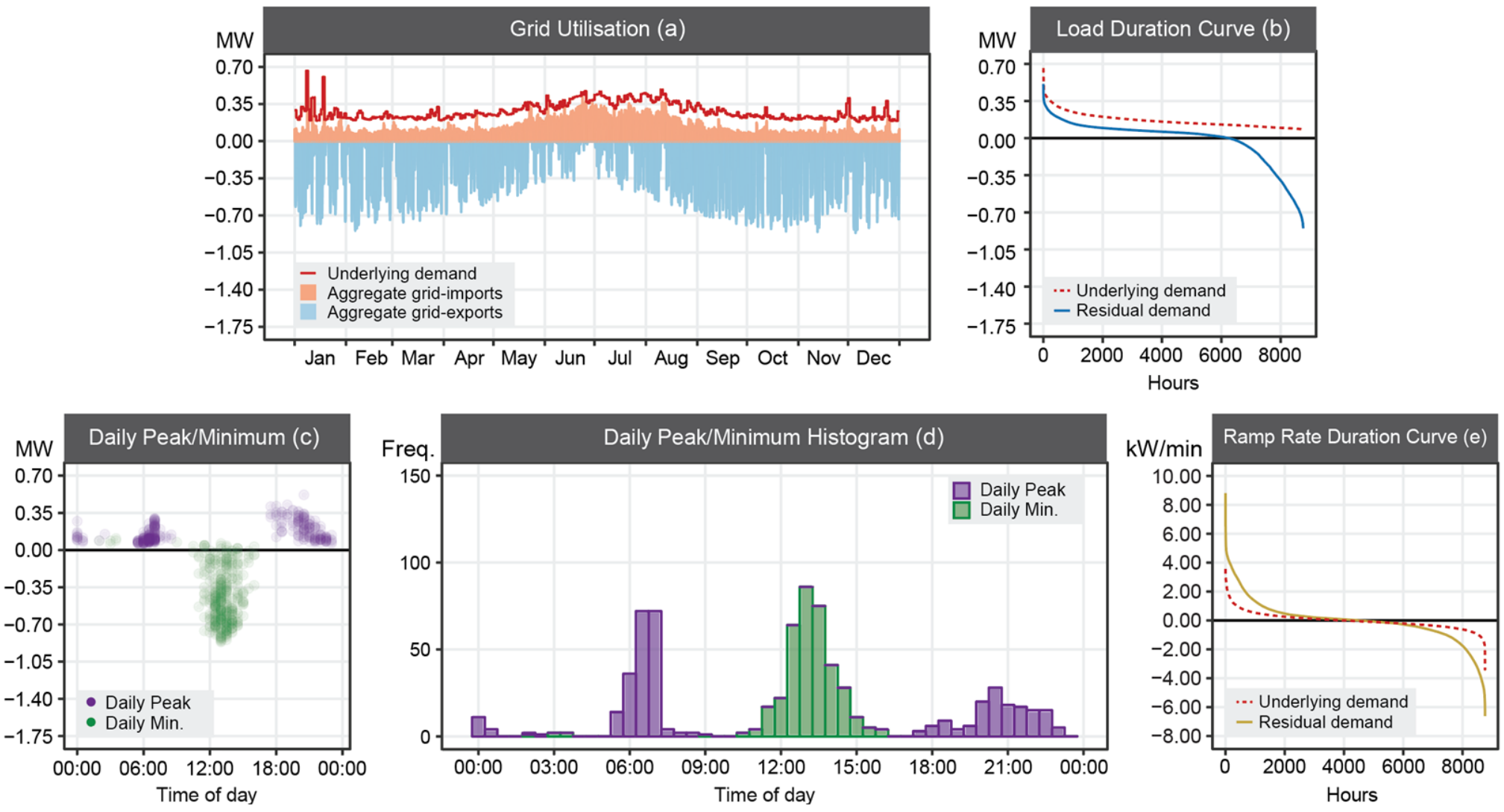

**Fig A.5.** Annual grid-utilisation (aggregate of 261 households) at the PV$_{M}$:B$_{S}$ (PV-medium and battery-small) operational stage. (a) Half-hourly grid-imports and grid-exports of the residual and underlying demand. (b) Load duration curve of residual and underlying demand. (c) Capacity and timing of diurnal demand peaks and minimums. (d) Histogram of diurnal demand peaks and minimums across each time interval. (e) Ramp rate duration curve of residual and underlying demand.

Publishing notes:
Display across the whole page.
Vector graphic provided in " Figure A.5.eps"

Compared to PV$_{S}$:B$_{S}$, the increase in installed PV capacity further reduces grid-imports across non-winter months (Fig. A.5a) with annual grid-imports falling to 559 MWh that peaks at 519 kW in the summer (Fig. A.5b). The increase in self-generation also leads to an increase in annual grid-exports to 701 MWh (Fig. A.5a) that peaks at 865 kW (Fig. A.5b). This grid-export peak exceeds the underlying peak demand of 663 kW, which has implications for network capacity constraints (discussed in Section 5.1.1). Moreover, annual grid-exports (701 MWh) exceeds annual grid-imports (559 MWh) leading to further grid and market implications (discussed in Section 5.2.1.1). The diurnal demand profile continues to shift, with peak demand in the evening between 20:00 and 23:00 diminishing, and occurring more consistently during the early-morning between 05:30 and 07:30 (Fig. A.5d). The diurnal minimum demand remains between 11:30 and 15:30 and continues to decrease in magnitude (Fig. A.5c).

## A.6. PV-medium and battery-medium ($PV_M$:$B_M$) stage

The $PV_M$:$B_M$ stage is represented with the time-series residual grid-utilisation from the $FiT_{25}$ scenario in the year 2030 (Table 3) with an average PV capacity of 5.96 $kW_P$ and an average battery capacity of 12.16 kWh per household (Fig. 4).

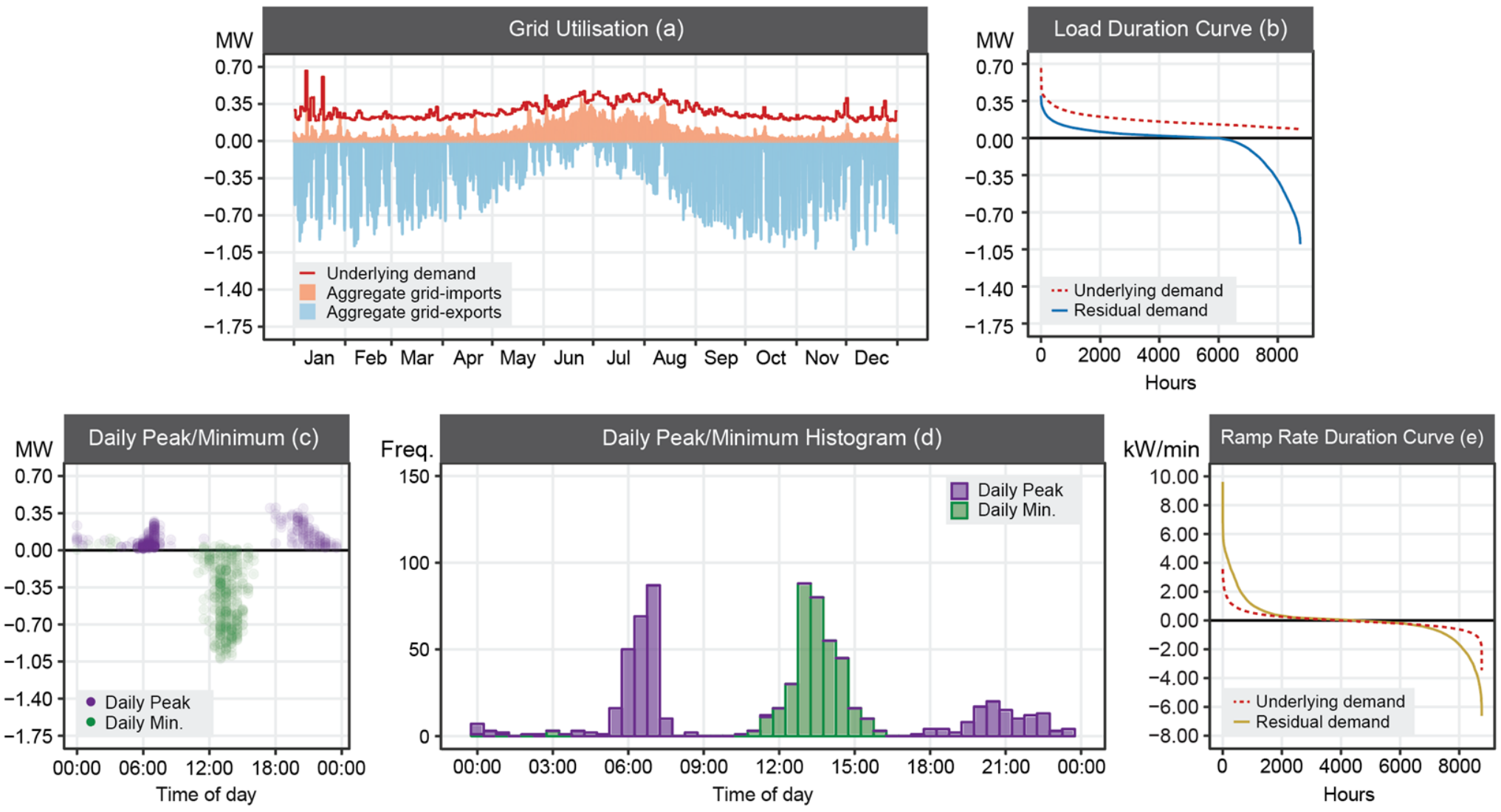

**Fig A.6.** Annual grid-utilisation (aggregate of 261 households) at the $PV_M$:$B_M$ (PV-medium and battery-medium) operational stage. (a) Half-hourly grid-imports and grid-exports of the residual and underlying demand. (b) Load duration curve of residual and underlying demand. (c) Capacity and timing of diurnal demand peaks and minimums. (d) Histogram of diurnal demand peaks and minimums across each time interval. (e) Ramp rate duration curve of residual and underlying demand.

Publishing notes:
Display across the whole page.
Vector graphic provided in " Figure A.6.eps"

Compared to $PV_M$:$B_S$, the additional installed battery capacity leads to further reductions in grid-imports in the non-winter months (Fig. A.6a). Annual grid-imports falls significantly to 340 MWh, which means only 23% of the underlying aggregate demand is supplied by the grid. Annual peak demand also falls to 400 kW (Fig. A.6b) and occurs in the winter (Fig. A.6a) indicating that network demand has become winter dominant (discussed in Section 5.1.3). Annual grid-exports increases slightly to 737 MWh (Fig. A.6a) with a higher peak of 1021 kW (Fig. A.6b). The diurnal demand profile also changes slightly, with the occurrence of the evening peak further diminishing between 20:00 and 23:00 and increasing in the early-morning between 05:30 and 08:00 (Fig. A.6d). The timing of diurnal minimum demand widens slightly to between 11:30 and 16:00 (Fig. A.6d).

### A.7. PV-medium and battery-large ($PV_M$:$B_L$) stage

The $PV_M$:$B_L$ stage is represented with the time-series residual grid-utilisation from the $FiT_{25}$ scenario in the year 2035 (Table 3) with an average PV capacity of 7.59 $kW_P$ and an average battery capacity 22.34 kWh per household (Fig. 4).

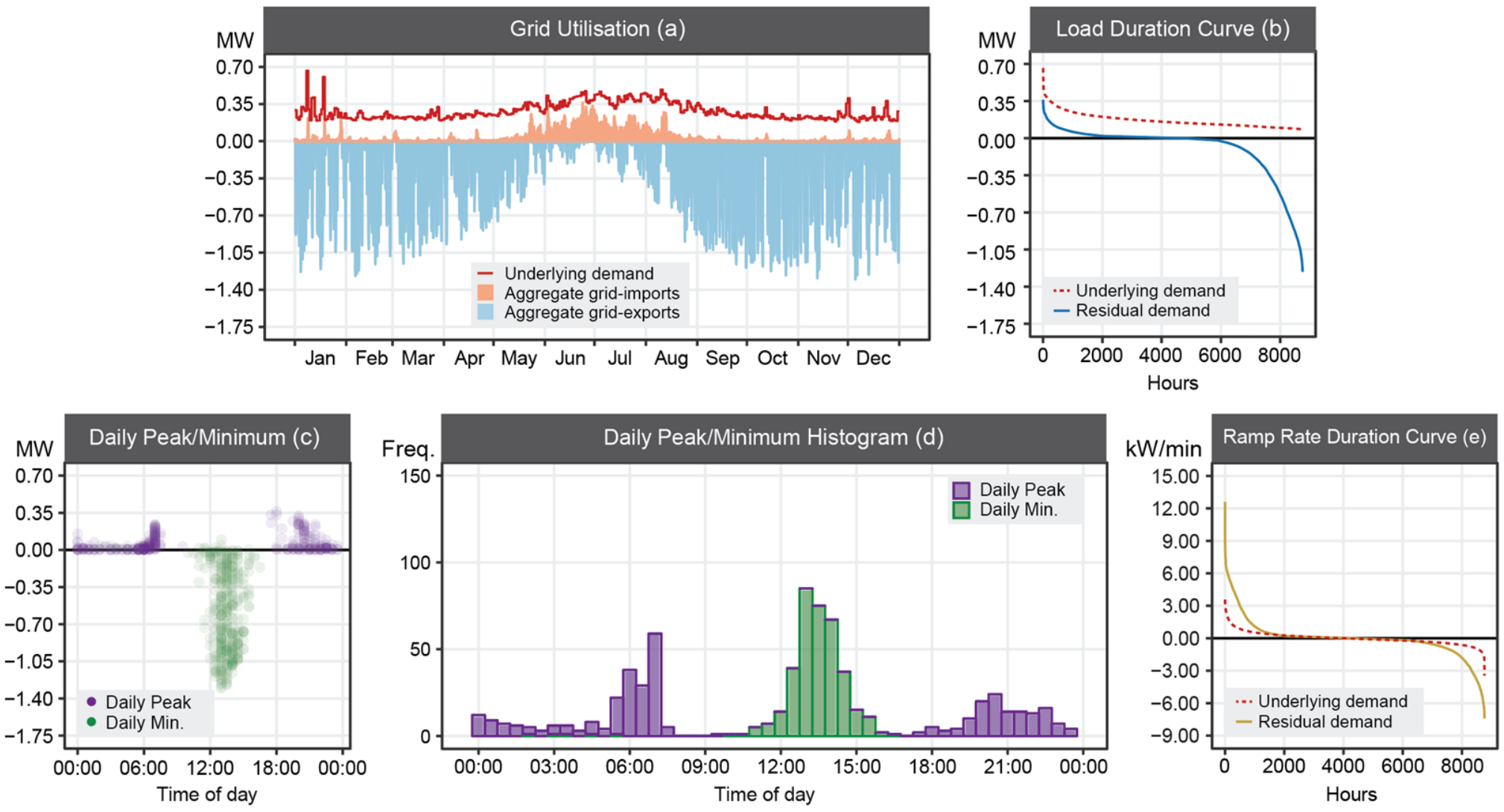

**Fig A.7.** Annual grid-utilisation (aggregate of 261 households) at the $PV_M$:$B_L$ (PV-medium and battery-large) operational stage. (a) Half-hourly grid-imports and grid-exports of the residual and underlying demand. (b) Load duration curve of residual and underlying demand. (c) Capacity and timing of diurnal demand peaks and minimums. (d) Histogram of diurnal demand peaks and minimums across each time interval. (e) Ramp rate duration curve of residual and underlying demand.

Publishing notes:
Display across the whole page.
Vector graphic provided in " Figure A.7.eps"

Compared to $PV_M$:$B_M$, the additional installed battery capacity leads to even further reductions in grid-imports across the non-winter months (Fig. A.7a) with annual grid-imports almost halving to 174 MWh, which means only 12% of the underlying aggregate demand is supplied by the grid. Annual peak demand reduces to 364 kW (Fig. A.7b) and continues to occur in the winter (Fig. 7a). Annual grid-exports increases to 1007 MWh (Fig. A.7a) that peaks at 1303 kW (Fig. A.7b). The timing of diurnal demand changes, with the early-morning peak between 05:30 and 07:30 reducing in occurrence, while the evening peak further widening to be between 19:30 and 00:30 (Fig. A.7d). The timing of diurnal minimum demand also shifts into the afternoon between 12:00 and 16:00 (Fig. A.7c and Fig. A.7d) as higher average battery capacities are able to store more self-generation.

## A.8. PV-large and battery-large ($PV_L$:$B_L$) stage

The $PV_L$:$B_L$ stage is represented with the time-series residual grid-utilisation from the $FiT_{25}$ scenario in the year 2037 (Table 3) with an average PV capacity of 8.30 $kW_P$ and an average battery capacity between 24.94 kWh per household (Fig. 4).

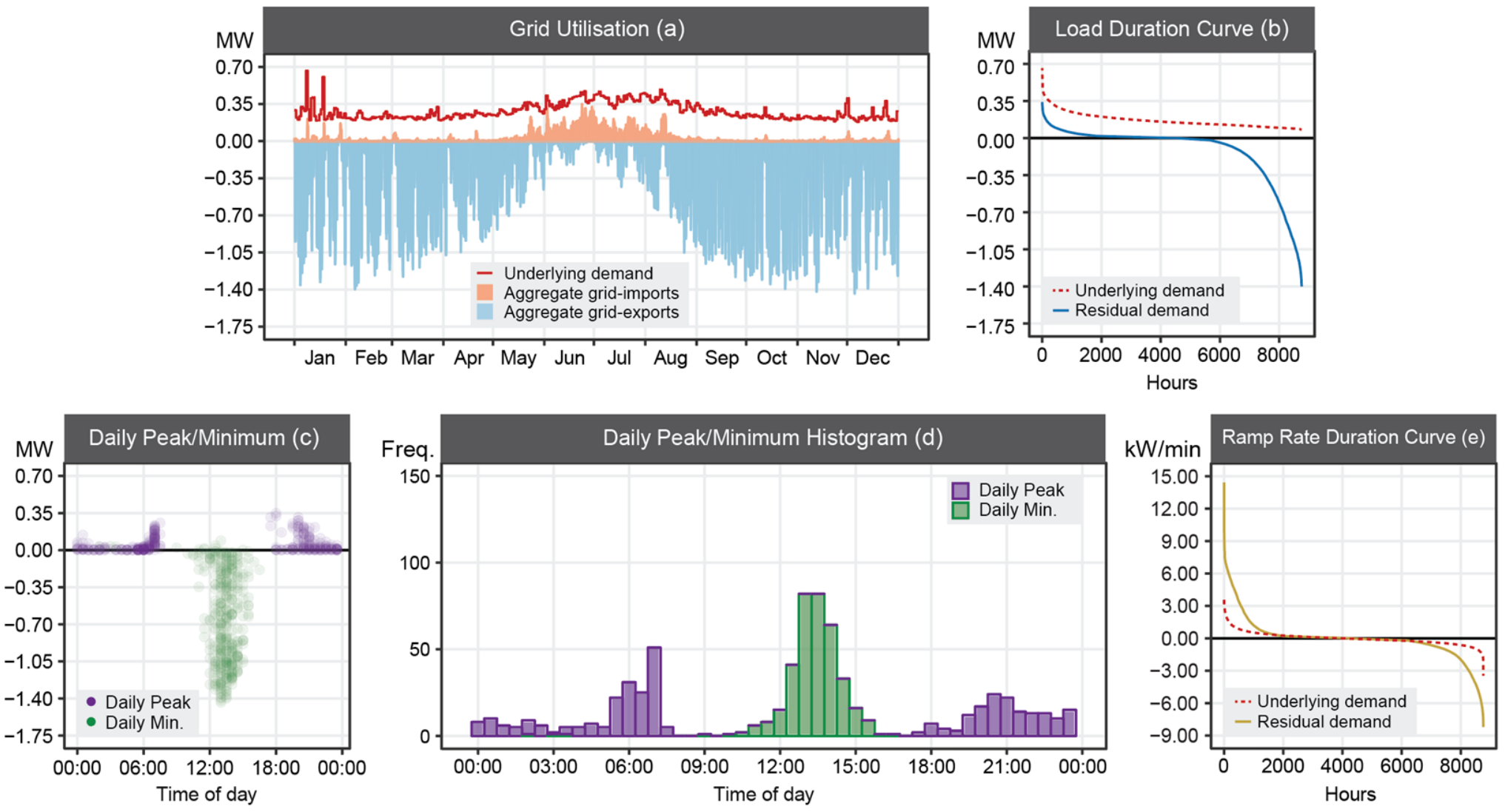

**Fig A.8** Annual grid-utilisation (aggregate of 261 households) at the $PV_L$:$B_L$ (PV-large and battery-large) operational stage. (a) Half-hourly grid-imports and grid-exports of the residual and underlying demand. (b) Load duration curve of residual and underlying demand. (c) Capacity and timing of diurnal demand peaks and minimums. (d) Histogram of diurnal demand peaks and minimums across each time interval. (e) Ramp rate duration curve of residual and underlying demand.

Publishing notes:
Display across the whole page.
Vector graphic provided in " Figure A.8.eps"

Compared to $PV_M$:$B_L$, the increased installed PV capacity further reduces annual grid-imports to 151 MWh that peaks at 349 kW in winter (Fig. A.8a and Fig. A.8b). Annual grid-exports continues to increase to 1166 MWh (Fig. 8a), which significantly exceeds annual grid-imports (discussed in Section 3.2). The grid-export peaks at 1438 kW (Fig. 8b) which is more than double the underlying demand peak of 663 kW. The timing of the diurnal demand mostly remains the same, with the early-morning peak remaining between 05:30 and 07:30 and the evening peak further widening to between 19:30 and 01:00 (Fig. A.8d). The timing of diurnal minimum demand remains in the afternoon between 12:00 and 15:30 (Fig. A.8c and Fig. A.8d).

## Appendix B. Sensitivity analysis

The low and high growth retail conditions assume the following parameters in Table B.1. All other parameters remain as defined in Table 1. In both sensitivity cases, the full range of FiT scenarios between 0% and 100% are evaluated. More detailed numerical results are included as part of the Research Data (Appendix D).

**Table B.1**
Additional low and high sensitivity cases with respect to the reference.

| Scenario parameter | Unit | Low | Reference | High |
|---|---|---|---|---|
| Discount rate | % / a | 10 | 6 | 2 |
| Change in tariff charges/rebates | % / a | 2 | 5 | 8 |
| Change in installed PV system costs | % / a | -3 | -5.9 | -9 |
| Change in installed battery system costs | % / a | -4 | -8 | -12 |

### B.1. Low growth case

Under the low growth case, the discount rate, projected reduction in PV and battery systems costs, retail tariffs and associated feed-in tariffs are reduced (Table B.1). This reduces the expected electricity bill cost savings while also raising upfront costs. Using the same capacity categories as from Table 2, the transition to PV-only and PV-only to PV-battery households occurs later while the average installed capacities are also reduced in each FiT scenario (Table

B.2). Furthermore, the transition from PV-only to PV-battery systems in the $FiT_{75}$ and $FiT_{100}$ scenarios are delayed beyond the 20-year evaluation period. However, the PV battery adoption pattern across each FiT scenario remained qualitatively similar when compared to the reference case (Table 3) but with a decelerated transition between each representative grid-operation stage. Numerical values are provided in the research data.

**Table B.2**
Grid-operation stages (based on the average PV and battery system capacities per household) for each FiT scenario in the low growth sensitivity scenario.

| Year | $FiT_0$ | $FiT_{25}$ | $FiT_{50}$ | $FiT_{75}$ | $FiT_{100}$ |
|---|---|---|---|---|---|
| 2018 | - | - | $PV_S$ | $PV_S$ | $PV_S$ |
| 2019 | - | - | $PV_S$ | $PV_S$ | $PV_S$ |
| 2020 | - | $PV_S$ | $PV_S$ | $PV_S$ | $PV_S$ |
| 2021 | $PV_S$ | $PV_S$ | $PV_S$ | $PV_S$ | $PV_M$ |
| 2022 | $PV_S$ | $PV_S$ | $PV_S$ | $PV_M$ | $PV_M$ |
| 2023 | $PV_S$ | $PV_S$ | $PV_S$ | $PV_M$ | $PV_M$ |
| 2024 | $PV_S$ | $PV_S$ | $PV_S$ | $PV_M$ | $PV_M$ |
| 2025 | $PV_S$ | $PV_S$ | $PV_S$ | $PV_M$ | $PV_M$ |
| 2026 | $PV_S$ | $PV_S$ | $PV_S$ | $PV_M$ | $PV_M$ |
| 2027 | $PV_S$ | $PV_S$ | $PV_S$ | $PV_M$ | $PV_M$ |
| 2028 | $PV_S$ | $PV_S$ | $PV_S$ | $PV_M$ | $PV_M$ |
| 2029 | $PV_S$ | $PV_S$ | $PV_M$ | $PV_M$ | $PV_M$ |
| 2030 | $PV_S$ | $PV_S$ | $PV_M$ | $PV_M$ | $PV_M$ |
| 2031 | $PV_S$ | $PV_S$ | $PV_M$ | $PV_M$ | $PV_M$ |
| 2032 | $PV_S$ | $PV_S$ | $PV_M$ | $PV_M$ | $PV_M$ |
| 2033 | $PV_S$ | $PV_S$ | $PV_M$ | $PV_M$ | $PV_M$ |
| 2034 | $PV_S$:$B_S$ | $PV_S$:$B_S$ | $PV_M$ | $PV_M$ | $PV_M$ |
| 2035 | $PV_S$:$B_S$ | $PV_S$:$B_S$ | $PV_M$ | $PV_M$ | $PV_M$ |
| 2036 | $PV_S$:$B_S$ | $PV_M$:$B_S$ | $PV_M$ | $PV_M$ | $PV_M$ |
| 2037 | $PV_S$:$B_S$ | $PV_M$:$B_S$ | $PV_M$ | $PV_M$ | $PV_M$ |

## B.2. High growth case

Under the high growth case, the discount rate, projected reduction in PV and battery systems costs, retail tariffs and associated feed-in tariffs are increased (Table B.1). This increases the expected electricity bill cost savings while also lowering upfront costs. As average PV and battery capacities in the later years exceeded the capacity categories from Table 2, PV capacities greater than 12 $kW_P$ and battery capacities greater than 30 kWh are respectively categorised as $PV_{XL}$ and $B_{XL}$. The transition from PV-only to PV-battery households occurs earlier while also increasing the average installed capacities in each FiT scenario (Table B.3). Furthermore, $FiT_{75}$ exhibited a transition pattern similar to $FiT_{50}$, where it quickly catches up to the average installed PV battery capacities in lower FiT scenarios. The PV battery adoption pattern across each FiT scenario remained qualitatively similar to the reference case (Table 3) but with an accelerated transition between each representative grid-operation stage. Numerical values are provided in the research data.

**Table B.3**
Grid-operation stages (based on the average PV and battery system capacities per household) for each FiT scenario in the high growth sensitivity scenario.

| Year | $FiT_0$ | $FiT_{25}$ | $FiT_{50}$ | $FiT_{75}$ | $FiT_{100}$ |
|---|---|---|---|---|---|
| 2018 | $PV_S$ | $PV_S$ | $PV_M$ | $PV_M$ | $PV_M$ |
| 2019 | $PV_S$ | $PV_S$ | $PV_M$ | $PV_M$ | $PV_M$ |
| 2020 | $PV_S$ | $PV_S$ | $PV_M$ | $PV_M$ | $PV_M$ |
| 2021 | $PV_S:B_S$ | $PV_S$ | $PV_M$ | $PV_M$ | $PV_M$ |
| 2022 | $PV_S:B_S$ | $PV_S$ | $PV_M$ | $PV_M$ | $PV_M$ |
| 2023 | $PV_M:B_S$ | $PV_M:B_S$ | $PV_M$ | $PV_M$ | $PV_M$ |
| 2024 | $PV_M:B_M$ | $PV_M:B_S$ | $PV_M$ | $PV_M$ | $PV_M$ |
| 2025 | $PV_M:B_M$ | $PV_M:B_S$ | $PV_M:B_S$ | $PV_M$ | $PV_M$ |
| 2026 | $PV_M:B_M$ | $PV_M:B_M$ | $PV_M:B_M$ | $PV_M:B_S$ | $PV_M$ |
| 2027 | $PV_M:B_L$ | $PV_M:B_M$ | $PV_M:B_M$ | $PV_M:B_S$ | $PV_M:B_S$ |
| 2028 | $PV_L:B_L$ | $PV_M:B_L$ | $PV_M:B_M$ | $PV_M:B_S$ | $PV_M:B_S$ |
| 2029 | $PV_L:B_L$ | $PV_M:B_L$ | $PV_M:B_M$ | $PV_M:B_S$ | $PV_M:B_S$ |
| 2030 | $PV_L:B_L$ | $PV_M:B_L$ | $PV_L:B_L$ | $PV_M:B_M$ | $PV_M:B_S$ |
| 2031 | $PV_L:B_{XL}$ | $PV_L:B_{XL}$ | $PV_L:B_L$ | $PV_L:B_M$ | $PV_M:B_S$ |
| 2032 | $PV_L:B_{XL}$ | $PV_L:B_{XL}$ | $PV_L:B_{XL}$ | $PV_L:B_L$ | $PV_M:B_S$ |
| 2033 | $PV_{XL}:B_{XL}$ | $PV_L:B_{XL}$ | $PV_L:B_{XL}$ | $PV_L:B_L$ | $PV_M:B_S$ |
| 2034 | $PV_{XL}:B_{XL}$ | $PV_L:B_{XL}$ | $PV_L:B_{XL}$ | $PV_L:B_L$ | $PV_L:B_M$ |
| 2035 | $PV_{XL}:B_{XL}$ | $PV_L:B_{XL}$ | $PV_{XL}:B_{XL}$ | $PV_{XL}:B_{XL}$ | $PV_L:B_M$ |
| 2036 | $PV_{XL}:B_{XL}$ | $PV_{XL}:B_{XL}$ | $PV_{XL}:B_{XL}$ | $PV_{XL}:B_{XL}$ | $PV_L:B_M$ |
| 2037 | $PV_{XL}:B_{XL}$ | $PV_{XL}:B_{XL}$ | $PV_{XL}:B_{XL}$ | $PV_{XL}:B_{XL}$ | $PV_L:B_L$ |

# Appendix C. Techno-economic model

## C.1. Key modelling assumptions

Electroscape is an open-source model written in R and developed as part of earlier research by the authors (Say et al., 2018, 2019). The aim of the model is to assess PV battery investment outcomes that are specific to a household's demand and insolation profile, expected retail and feed-in tariffs, and installed PV battery system costs. Electroscape consists of three components, a *technical model* that evaluates the operation of a specific PV and/or battery capacity on a household's load profile, a *financial model* that calculates the financial viability of that choice, and an *investment decision model* that evaluates across a range of PV battery combinations to determine if the household should make an investment, and the most suitable PV and/or battery capacity to invest in. If an investment is made, the household demand profile is updated, and the process repeats for the following year. The grid-utilisation changes are generated by applying Electroscape to each of the 261 households and aggregating the results. The key modelling assumptions in this paper are summarised as follows:

- Underlying household demand profiles, insolation profiles, expected retail and feed-in tariffs, and expected PV battery system costs are exogenous parameters.
- Feed-in tariff payments are only eligible for households with a combined PV capacity of 5 $kW_P$ and under.
- Household demand and insolation profiles from Sydney, Australia are used as representative examples for Perth, Australia.
- The meter data from Sydney, Australia consists of half-hourly resolution underlying demand and PV generation data between 1st July 2012 and 30th June 2013 collected from gross utility energy meters (Ausgrid, 2018).

- Insolation profiles are calculated by normalising the PV generation utility meter data with the declared PV capacity.
- The annual underlying demand and insolation profiles for each household are repeated for each year of the simulation.
- The PV battery investment transition results for each household are independently evaluated.
- PV generation performance degrades linearly.
- Battery energy storage capacity degrades linearly.
- Battery charge/discharge efficiencies and operational performance remain constant over its operational lifespan.
- The operational lifespan of battery systems matches their warranty period of 10 years.
- Batteries operate to maximise PV self-consumption and with flat tariffs, grid-charging and grid-discharging operation is not evaluated.

## C.2. Technical model

For a single household, the *investment decision model* requires the discounted cash flows over a 10-year financial horizon from potentially installing a range of PV battery combinations. The *technical model* provides the 10-year grid-utilisation profile used to define these discounted cash flows. Given a specific PV capacity ($p$) and battery capacity ($b$) the *technical model* will simulate their operation using a household's load and insolation profiles (at 30-min resolution). The PV generation profile is calculated by scaling the insolation profile by the PV capacity with a linear degradation in generation (80% capacity after 25-years). An 'intermediate net-load' profile is calculated by subtracting the PV generation profile from the household's load profile. A battery simulation model (based on the Tesla Powerwall 2) maximises PV self-consumption by using excess generation from the 'intermediate net-load' to charge the battery (while under a 5-kW limit) until it is full, and during times where PV generation is below that of 'intermediate net-load', the battery will discharge (while under the 5 kW limit) until it is empty. Reflecting technical specifications, the battery simulation model assumes a 100% depth-of-discharge, round-trip efficiency of 89%, and a linear degradation in battery capacity (70% remaining after 10-years). After the battery simulation model, the resulting residual load profile reflects the grid-utilisation after having installed the specific PV and battery system.

## C.3. Financial and investment decision model

The residual load profile coupled with the projected retail tariffs and FiT is used to determine the expected electricity bills over the next 10 years. By comparing this electricity bill with a scenario without additional PV or battery systems installed, the expected cash flow over 10-years are calculated (C.2). By factoring in the installation cost of the PV ($p$) and battery ($b$) system (C.8) and the discount rate ($R_d$), the value of the PV battery investment can be expressed as NPV (C.1). Each PV battery combination is then evaluated and treated as competing investment opportunities and valued according to their NPV (C.9). The PV battery configuration with the largest NPV becomes a prime candidate for installation.

To represent a minimum level of awareness and investment confidence before households commit their limited financial capital, an additional test is performed. Reflecting the use of discounted payback periods by the AEC (2019) to report on the attractiveness of PV investments across Australia, this model requires at least one of the evaluated PV battery configurations to have a discounted payback of under 5 years before deciding to install the PV and/or battery system with the highest NPV. Once the system is installed, the household load profile is updated, and all subsequent PV battery investments must consider this installed system. This allows the model to simulate how households transition to different types of PV battery investments as the retail cost factors change over time. The PV battery investment results for each household are provided in the research data.

As previously described in Say et al. (2019) the financial equations are as follows, the profitability of each PV and battery investment in each year (*t*) of the simulation can be expressed as an NPV that depends on discounted annual cash flows over the 10-year investment horizon (*N*) and upfront system costs.

$$NPV(p,b,t) = \sum_{n=1}^{10} \frac{Cash\ Flow(p,b,n,t)}{(1+R_d)^n} - Cost(p,b,t) \qquad \text{(C.1)}$$

where, $p$ = Rated PV capacity (kWP)

$b$ = Battery energy storage capacity (kWh)

$$Cash\ Flow(p,b,n,t) = Bills_{Base}(n,t) - Bills_{System}(p,b,n,t) \qquad \text{(C.2)}$$

where, *Base* is the cost of electricity without the proposed PV or battery system; and

*System* is the cost of electricity with a particular PV battery system

The *Base* and *System* electricity costs for each *n*-th year from the *t*-th forecast year are given by:

$$Bills_{Base}(n,t) = E_{Import}(0,0,n) \cdot T_{Import}(n,t) - E_{Export}(0,0,n) \cdot T_{Export}(0,n,t) + 365 \cdot T_{Daily}(n,t) \qquad \text{(C.3)}$$

$$Bills_{System}(p,b,n,t) = E_{Import}(p,b,n) \cdot T_{Import}(n,t) - E_{Export}(p,b,n) \cdot T_{Export}(p,n,t) + \ + 365 \cdot T\ \ (n,t) \qquad \text{(C.4)}$$

where, $T_{Import}(n,t) = T_{Import_Start} \cdot \left(1 + R_{Tariffs}\right)^{n+t-2}$ (C.5)

$$T_{Export}(p,n,t) = \begin{cases} T_{Export_Start} \cdot \left(1 + R_{Tariffs}\right)^{n+t-2}, & p \leq P_{Export_Limit} \\ 0, & otherwise \end{cases} \qquad \text{(C.6)}$$

$$T_{Daily}(n,t) = T_{Daily_Start} \cdot \left(1 + R_{Tariffs}\right)^{n+t-2} \qquad \text{(C.7)}$$

The system cost is given by:

$$Cost(p,b,t) = p \cdot C_{PV_Start} \cdot (1 + R_{PV})^{t-1} + b \cdot C_{Battery_Start} \cdot \left(1 + R_{Battery}\right)^{t-1} \qquad \text{(C.8)}$$

The PV and battery configuration with the highest NPV in the $t$-th year is chosen according to the equation following, with the PV and battery capacities defined as $p_c$ and $b_c$ respectively:

$$Investment\ Decision(t) = Max\ [NPV(p, b, t)] \tag{C.9}$$

$$\text{where,}\quad 0 \leq p \leq P^* \ and\ P'' = \begin{cases} 10\ \text{kW}_\text{P} & \text{, initial} \\ P' \cdot (1 + 40\%), & \text{if } Investment\ Choice(t) = (P', b_c) \\ P' & \text{, otherwise} \end{cases} \tag{C.10}$$

$$0 \leq b \leq B^* \ and\ B'' = \begin{cases} 20\ \text{kWh} & \text{, initial} \\ B' \cdot (1 + 40\%), & \text{if } Investment\ Choice(t) = (p_c, B') \\ B' & \text{, otherwise} \end{cases} \tag{C.11}$$

The decision to invest in a given year (t) depends on the Discounted Payback Period (DPP) for one of the assessed PV (p) and battery (b) combinations to be less than or equal to 5 years.

$$DPP[-Cost(p, b, t), Cashflow(p, b, n, t)] \leq 5 \tag{C.12}$$

with $n = \{1, 2, \ldots, 10\}$

# Appendix D. Research data

The R open-source code, demand profile data, insolation data, household investment analysis, sensitivity and computational results are publicly accessible from https://doi.org/10.25917/5ea7d261ae32a.